\documentclass[twocolumn]{aastex63}

\usepackage{graphicx}
\usepackage{advdate}
\usepackage{amsmath}

\newcommand{\asec}{\hbox to 1pt{}\rlap{$^{\prime\prime}$}.\hbox to 2pt{}}
\newcommand{\amin}{\hbox to 1pt{}\rlap{$^{\prime}$}.\hbox to 1pt{}}
\newcommand{\adeg}{\hbox to 1pt{}\rlap{$^{\circ}$}.\hbox to 2pt{}}

\accepted{for publication in {\it The Astrophysical Journal}}
\shortauthors{Lauer, Postman, Weaver et al.}
\shorttitle{Cosmic Optical Background}

\begin{document}

\title{New Horizons Observations of the Cosmic Optical Background}

\author{Tod R. Lauer}
\affil{NSF's National Optical Infrared Astronomy Research
Laboratory,\footnote{The NSF's OIR Lab is operated by AURA, Inc.
under cooperative agreement with NSF.}
P.O. Box 26732, Tucson, AZ 85726}

\author{Marc Postman}
\affil{Space Telescope Science Institute,\footnote{Operated by AURA, Inc., for the National Aeronautics and Space Administration}
3700 San Martin Drive, Baltimore, MD 21218}

\author{Harold A. Weaver}
\affil{The Johns Hopkins University Applied Physics Laboratory,
Laurel, MD 20723-6099}

\author{John R. Spencer}
\affil{Department of Space Studies, Southwest Research Institute, 1050 Walnut St., Suite 300, Boulder, CO 80302}

\author{S. Alan Stern}
\affil{Space Science and Engineering Division, Southwest Research Institute, 1050 Walnut St., Suite 300, Boulder, CO 80302}

\author{Marc W. Buie}
\affil{Department of Space Studies, Southwest Research Institute, 1050 Walnut St., Suite 300, Boulder, CO 80302}

\author{Daniel D. Durda}
\affil{Department of Space Studies, Southwest Research Institute, 1050 Walnut St., Suite 300, Boulder, CO 80302}

\author{Carey  M. Lisse}
\affil{The Johns Hopkins University Applied Physics Laboratory,
Laurel, MD 20723-6099}

\author{A. R. Poppe}
\affil{Space Sciences Laboratory, University of California at Berkeley, 7 Gauss Way, Berkeley, CA 94720, USA}

\author{Richard P. Binzel}
\affil{Department of Earth, Atmospheric, and Planetary Sciences, Massachusetts Institute of Technology, Cambridge, MA 02139}

\author{Daniel T. Britt}
\affil{Department of Physics, University of Central Florida, Orlando, FL 32816}

\author{Bonnie J. Buratti}
\affil{Jet Propulsion Laboratory, California Institute of Technology, Pasadena, CA 91109}

\author{Andrew F. Cheng}
\affil{The Johns Hopkins University Applied Physics Laboratory,
Laurel, MD 20723-6099}

\author{W.M. Grundy}
\affil{Lowell Observatory, Flagstaff, AZ 86001}

\author{Mihaly Hor\'{a}nyi}
\affil{Laboratory for Atmospheric and Space Physics, University of Colorado, Boulder, CO 80303}

\author{J.J. Kavelaars}
\affil{National Research Council of Canada, Victoria BC \& Department of Physics and Astronomy, University of Victoria, Victoria, BC}

\author{Ivan R. Linscott}
\affil{Independent consultant, Mountain View, CA 94043}

\author{William B. McKinnon}
\affil{Dept. of Earth and Planetary Sciences and McDonnell Center for the Space Sciences, Washington University, St. Louis, MO 63130}

\author{Jeffrey M. Moore}
\affil{NASA Ames Research Center, Space Science Division, Moffett Field, CA 94035}

\author{J. I. N\'{u}\~{n}ez}
\affil{The Johns Hopkins University Applied Physics Laboratory,
Laurel, MD 20723-6099}

\author{Catherine B. Olkin}
\affil{Department of Space Studies, Southwest Research Institute, 1050 Walnut St., Suite 300, Boulder, CO 80302}

\author{Joel W. Parker}
\affil{Department of Space Studies, Southwest Research Institute, 1050 Walnut St., Suite 300, Boulder, CO 80302}

\author{Simon B. Porter}
\affil{Department of Space Studies, Southwest Research Institute, 1050 Walnut St., Suite 300, Boulder, CO 80302}

\author{Dennis C. Reuter}
\affil{NASA Goddard Space Flight Center, Greenbelt, MD 20771}
;
\author{Stuart J. Robbins}
\affil{Department of Space Studies, Southwest Research Institute, 1050 Walnut St., Suite 300, Boulder, CO 80302}

\author{Paul Schenk}
\affil{Lunar and Planetary Institute, Houston, TX 77058}

\author{Mark R. Showalter}
\affil{SETI Institute, Mountain View, CA 94043}

\author{Kelsi N. Singer}
\affil{Department of Space Studies, Southwest Research Institute, 1050 Walnut St., Suite 300, Boulder, CO 80302}

\author{Anne. J. Verbiscer}
\affil{University of Virginia, Charlottesville, VA 22904}

\author{Leslie A. Young}
\affil{Department of Space Studies, Southwest Research Institute, 1050 Walnut St., Suite 300, Boulder, CO 80302}

\begin{abstract}

We used existing data from the New Horizons LORRI camera to measure the optical-band ($0.4\lesssim\lambda\lesssim0.9{\rm\mu m}$) sky brightness within seven high galactic latitude fields. The average raw level measured while New Horizons was 42 to 45 AU from the Sun is $33.2\pm0.5{\rm ~nW ~m^{-2} ~sr^{-1}}.$ This is $\sim10\times$ darker than the darkest sky accessible to the {\it Hubble Space Telescope}, highlighting the utility of New Horizons for detecting the cosmic optical background (COB). Isolating the COB contribution to the raw total requires subtracting scattered light from bright stars and galaxies, faint stars below the photometric detection-limit within the fields, and diffuse Milky Way light scattered by infrared cirrus. We remove newly identified residual zodiacal light from the IRIS $100\mu$m all sky maps to generate two different estimates for the diffuse galactic light (DGL). Using these yields a highly significant detection of the COB in the range ${\rm 15.9\pm 4.2\ (1.8~stat., 3.7~sys.) ~nW ~m^{-2} ~sr^{-1}}$ to ${\rm 18.7\pm 3.8\ (1.8~stat., 3.3 ~sys.)~ nW ~m^{-2} ~sr^{-1}}$ at the LORRI pivot wavelength of 0.608 $\mu$m. Subtraction of the integrated light of galaxies (IGL) fainter than the photometric detection-limit from the total COB level leaves a diffuse flux component of unknown origin in the range ${\rm 8.8\pm4.9\ (1.8 ~stat., 4.5 ~sys.) ~nW ~m^{-2} ~sr^{-1}}$ to ${\rm 11.9\pm4.6\ (1.8 ~stat., 4.2 ~sys.) ~nW ~m^{-2} ~sr^{-1}}$. Explaining it with undetected galaxies requires the galaxy-count faint-end slope to steepen markedly at $V>24$ or that existing surveys are missing half the galaxies with $V< 30$.

\end{abstract}


\section{How Dark Does the Sky Get?}

The simple fact that it's dark at night is known as ``Olber's paradox," and argues that the Universe is finite in time or space \citep{olber}. Further insight into the formation and evolution of the Universe comes from asking exactly how dark the night sky is. The cosmic optical background (COB) is the average flux of visible light photons averaged over the volume of the observable Universe. It reflects, at least in part, an integral over the cosmological history of star formation  occurring in recognizable galaxies, proto-galaxies, and star clusters \citep{conselice}, as well as mass accretion by black holes (BHs) associated with the systems.  A diffuse component of the COB {\it not} associated with any presently recognizable objects may determine how much star formation and active galactic nuclei power comes from stars or black holes in nominally low-density regions of the Universe or that formed prior to the organization of stars and BHs into recognizable associations. A diffuse COB  component (dCOB) may also reflect the more exotic production of photons by the annihilation or decay of dark matter particles \citep[e.g.][]{maurer, gong}. Detection of a genuine dCOB, however, has remained elusive despite several attempts  to search for it \citep{cooray}.  Our goal is to measure the flux associated with the  COB and to test for evidence of a dCOB.

Observation of the COB is challenging, as the total sky level measured by an astronomical instrument is the integral over several components that are stronger than the COB. Rich knowledge and accurate calibration of the low light-level performance of the instrument, itself, is also required.   Presently, all techniques to measure the optical sky (directly) are done with spacecraft to avoid the effects of airglow and artificial light on ground-based measurements. However, observations made from the inner solar system are still strongly dominated by zodiacal light (ZL). The darkest sky available to the {\it Hubble Space Telescope}, for example, is at the north ecliptic pole, but even there ZL is still over an order of magnitude stronger than limits on the COB discussed in the literature.

\begin{figure*}[htbp]
\centering
\includegraphics[keepaspectratio,width=6.0 in]{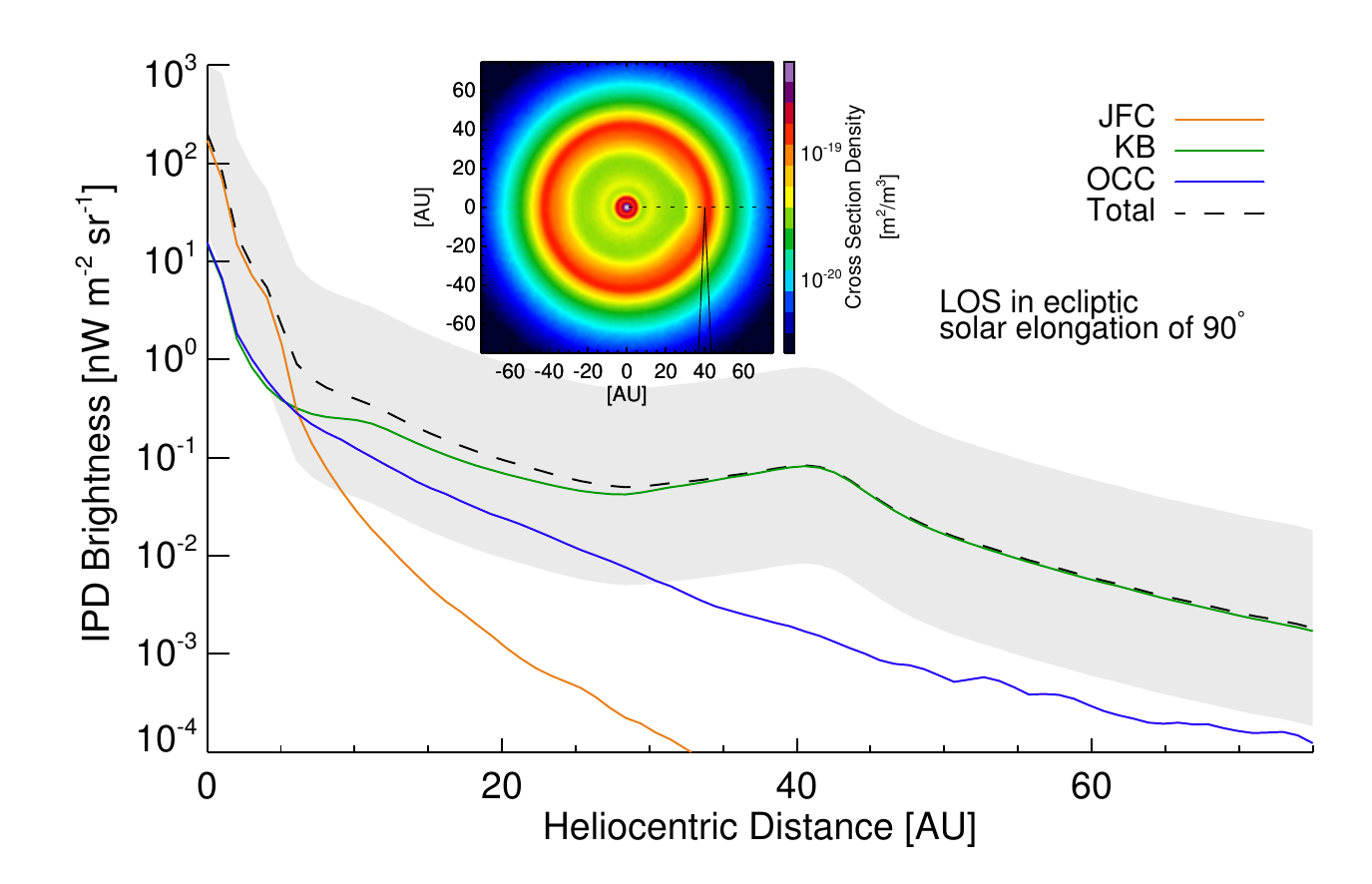}
\caption{A model of the expected flux density of sunlight scattered by interplanetary dust (IPD) over the outer solar system, as seen with a $90^\circ$ solar elongation \citep{zem18, poppe} line-of-sight (LOS). The sources of dust included in the model are Jupiter-family comets (JFC), collisions in the Kuiper belt (KB), and Oort-cloud comets (OCC). The inset figure gives dust cross-section density over the ecliptic plane.}
\label{fig:dust}
\end{figure*}

In contrast, ZL appears to be negligible in the {\it outer} solar system. Figure \ref{fig:dust} shows the best estimate of the flux of sunlight scattered by interplanetary dust (IPD) as a function of distance from the Sun. This will be discussed in detail in $\S\ref{sec:ipd},$ but in short, at distances beyond 10 AU, the estimated ZL flux is well below the expected COB flux.  The utility of COB measures from an outer solar system vantage point motivated \citet{zemcov} to use archival images obtained by NASA's New Horizons spacecraft during its cruise to Pluto to derive upper limits on the COB.  From the limited observations available, \citet{zemcov} derived COB flux upper limits compatible with previous work.  They also argued that a dedicated New Horizons program of sky observations might substantially improve constraints on the COB.

New Horizons is presently travelling out of the solar system at over 14 km/s. After its encounter with Pluto on July 14, 2015 \citep{pluto} at 32.9 AU from the Sun, it has been traversing the Kuiper Belt, encountering the Kuiper Belt object (KBO) Arrokoth on January 1, 2019 \citep{arrokoth} at 43.3 AU from the Sun. Over the last few years New Horizons has also been used to image distant unresolved KBOs (DKBOs) to search for satellites, obtain light curves, and determine their phase coefficients \citep{dkbo1, dkbo2}. A side product of this program has been the serendipitous sampling of the background sky at large distances from the Sun. 

There are now several New Horizons imaging data sets obtained at $>40$ AU from the Sun, with fields having large solar elongations and high galactic latitudes. Improvements in the operation of the LORRI camera on board New Horizons provide for longer and hence deeper exposures than were possible prior to the Pluto encounter.  Continued analysis of LORRI's performance has provided for more accurate photometry at low signal levels.  All of these factors motivate a new attempt to constrain the COB and search for a dCOB.  We discuss archival LORRI data-sets suitable for the detection of faint sky signals in $\S\ref{sec:obs},$ and the reduction of the images and measurement of total-sky levels in $\S\ref{sec:measure}.$ Decomposition of the sky signals to correct for known light sources is presented in $\S\ref{sec:decomp},$ and review of the resultant COB signal and comparison of it to previous measurements is presented in $\S\ref{sec:detect}.$

\section{Dark Sky Image-Sets}\label{sec:obs}

\begin{figure*}[htbp]
\centering
\includegraphics[keepaspectratio,width=6 in]{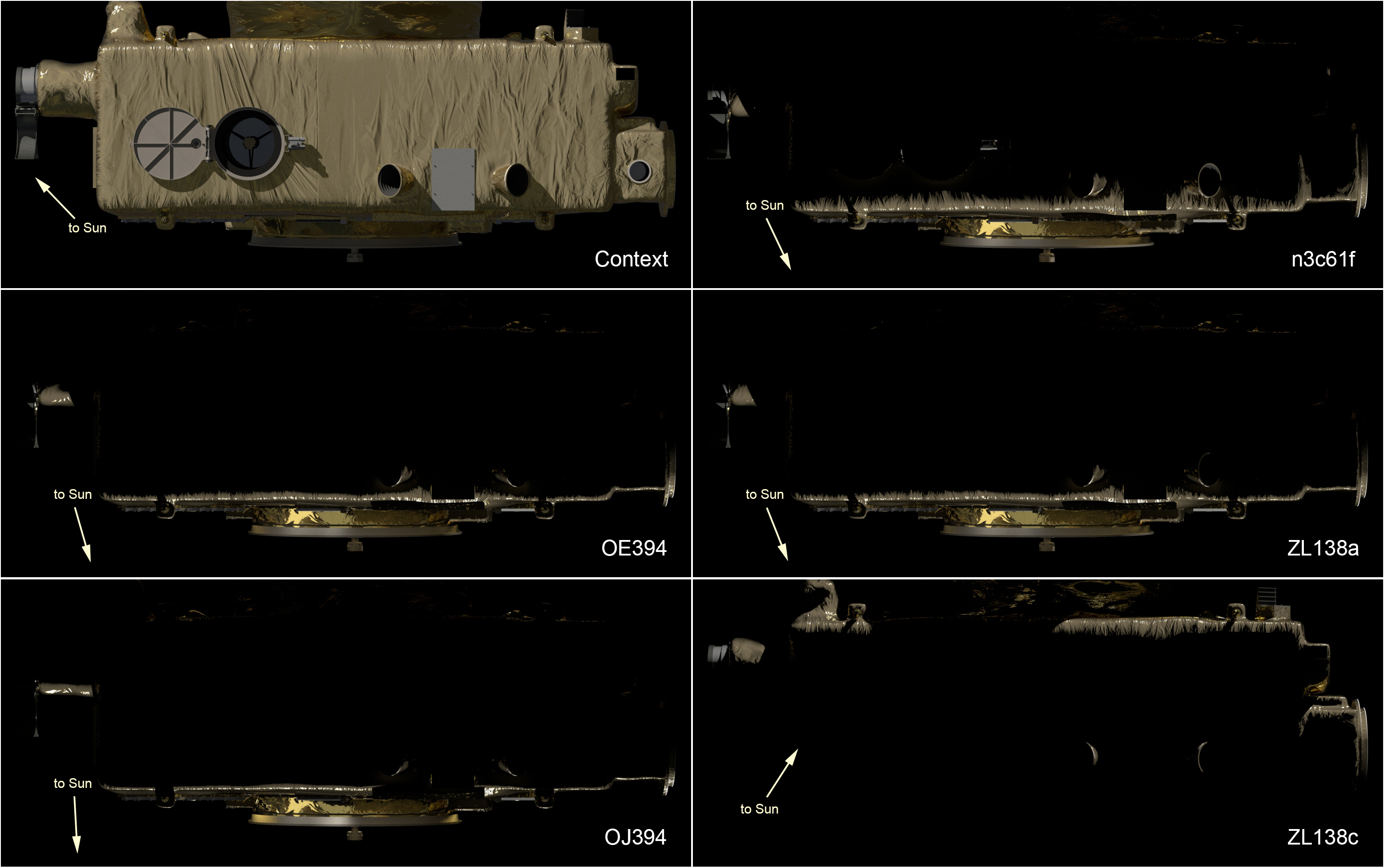}
\caption{The top left panel shows a rendering of the ``LORRI-side" panel on the New Horizons spacecraft.  For orientation, the spacecraft's  high gain dish-antenna is out of the picture to the top. The spacecraft's  distinctive radioisotope thermoelectric generator is on the opposite side of the spacecraft from LORRI. In this rendering the LORRI aperture cover has been permanently opened and is stowed to the left of the  LORRI aperture.  The spacecraft's two star trackers are mounted on the right side of this panel. The other five renderings show the geometry of the solar illumination of the LORRI-side for the seven sky fields (the  ZL138b field has essentially the same geometry as ZL138a,  and ZL138d as that  of ZL138c, and are not shown). LORRI is in  the spacecraft shadow, but some sunlight still reaches the star trackers at all pointings. The ray-tracing software shows that no significant flux of scattered sunlight enters the LORRI aperture, however.}
\label{fig:lorri_side}
\end{figure*}

The bulk of the deep imaging observations conducted by New Horizons has been done using its LORRI (the Long-Range Reconnaissance Imager) instrument. Complete details on LORRI are provided by \citet{lorri} and \citet{lorri2}, but in brief, it is an unfiltered (white light) $1032\times1024$ pixel CCD imager mounted on a 20.9 cm aperture Cassegrain reflector. In detail, the telescope itself is mounted in the  interior of the main spacecraft bus and looks out an aperture in one of the bus's side panels (Figure \ref{fig:lorri_side}). An aperture door protected the telescope during launch and the initial phases of the mission, but was opened permanently prior to the Jupiter flyby in 2007.

Following 1024 columns of pixels in the active imaging area, there are eight columns of unilluminated pixels. This portion of the sensor is covered with a metal strip; the shielded pixels in a given row are read out last. The first four of these are discarded, but the final four columns are used to measure the bias level and any integrated dark current over the duration of an exposure. 

\begin{figure}[htbp]
\centering
\includegraphics[keepaspectratio,width=3.5 in]{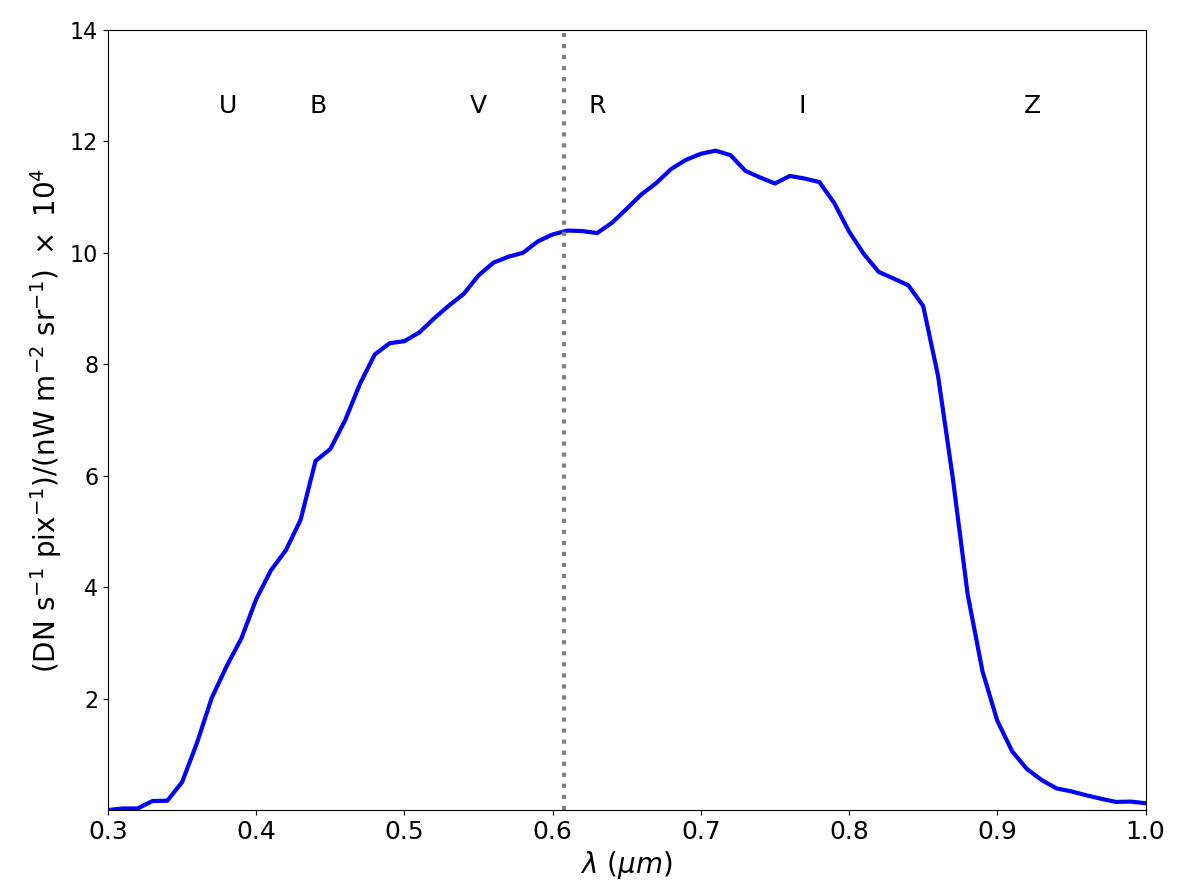}
\caption{The absolute responsivity function of the LORRI instrument in the $4\times4$ pixel binning mode. This figure is reproduced from Figure 24 in \citet{lorri2}. The vertical dashed line denotes the pivot wavelength of 0.608$~\mu$m, derived from the LORRI QE curve. The central wavelengths of common broadband photometric filters are shown for reference.}
\label{fig:lorribandpass}
\end{figure}

For deep observations, the camera is operated with $4\times4$ pixel binning, producing (raw) images in $257\times256$ pixel format with a single bias/dark column. The pixel-scale in this mode is $4\asec08,$ which provides a $17\amin4$ field. LORRI's sensitivity extends from the blue to NIR and is defined by the CCD response and telescope optics (Figure~\ref{fig:lorribandpass}).  The pivot wavelength is 0.608$~\mu$m.  The camera is operated with a gain of $19.4e^-$ per 1 DN (data number), and the read-noise is $24e^-.$ In $4\times4$ mode the photometric zeropoint is $18.88\pm0.01$ AB magnitudes corresponding to a 1 DN/s exposure level \citep{lorri2}. 

While all LORRI images, except those taken during the closest approach to Pluto, will have astronomical sky in some portion of the field, in practice most LORRI programs will not be suitable for measurement of the faint background sky.  Both Pluto and Arrokoth were projected against the center of the Milky Way during the approach phase, and were angularly close to the Sun during the departure phase.  To provide for the most sensitive measurements of the sky, we selected previously existing LORRI data sets that satisfied the following criteria:

\begin{itemize}
\item The field was observed at a solar elongation
greater than $90^\circ.$
\item The field had a galactic latitude
$|b|\ge50^\circ.$
\item The image exposure time was 30s or greater.
\item The dataset sequences had to run for at least 150s after the start of the first exposure in the sequence to avoid an initially elevated background.
\end{itemize}

 The galactic latitude limit is somewhat arbitrary; however, at small latitudes the density of stars and the diffuse galaxy light scattered off the infra-red cirrus are likely to overwhelm the COB. The limit on image exposure time is to provide for the best separation of the sky signal from complex random structure in the LORRI bias signal.
 
 \subsection{The New Horizons Shadow}
 
 Scattered sunlight contributes significantly to the LORRI background at solar elongation angles ${\rm (SEA) <90^\circ.}$ Many DKBO observations were taken in such ``bright" conditions at Sun angles well interior to this limit, but there is no calibration methodology that would cleanly isolate the astronomical sky from their considerably stronger scattered light backgrounds.  
 
 For ${\rm  SEA >90^\circ,}$ LORRI is in the shadow of the spacecraft, so sunlight  can no longer enter its aperture {\it directly}. However, spacecraft components extending out from the LORRI-side panel, such as the star trackers, may still be in sunlight, depending on the particular geometry of the observation, and thus may scatter sunlight {\it indirectly} into the LORRI aperture at a shallow angle.  This problem becomes less important as the SEA increases past $90^\circ$, but because the star trackers are mounted close to the bottom edge of the  LORRI-side panel (see Figure \ref{fig:lorri_side}), surprisingly large SEAs are required, if the sunlight is  coming from below  the spacecraft, to have them completely shadowed.  At the same time, because the New Horizon trajectory  out of  the solar system is directed towards the galactic center, SEAs approaching $180^\circ$ will always have low galactic latitudes.  This means that using LORRI to measure the COB at high-galactic  latitudes requires evaluating the effects of scattered sunlight for the likely SEAs, even  when the LORRI aperture itself is shadowed.
 
 To explore the importance of indirect sunlight for our selected fields, we used ray-tracing on a model of the spacecraft generated by one of the authors (DDD).  The model was developed for use in artistic renditions and is faithful to the geometric configuration of the various spacecraft components, while making plausible  assumptions about their surface properties and reflectivities. We thus consider the renderings useful for an overall  investigation into whether or not scattered sunlight might be important at the order-of-magnitude level, while being wary of using it for highly precise corrections.
 
 With these caveats, we generated illuminated renditions of the full spacecraft appropriate to the spacecraft attitudes used to observe the fields discussed at the end of this section\footnote{The fields of the zodiacal light sequences were obtained in two pairs of fields with only small angular separations, so we present only five rather than seven renderings.}. The LORRI-side panels are shown in Figure  \ref{fig:lorri_side}.  As can be seen, the two star trackers do remain slightly illuminated for all fields, creating glints that can be observed from the entrance of the LORRI aperture.
 
 Evaluation of the effect of the scattered sunlight was done  by filling the aperture with a white screen and measuring its average illumination from the star tracker glints.  To calibrate this  measure we then illuminated the screen with direct sunlight at the elevation angle and direction of the star tracker glints as seen at the aperture entrance.  From DKBO images obtained at $70^\circ<{\rm SEA}< 90^\circ,$ we know the sky level generated by direct sunlight as a function of its incidence angle with respect to the LORRI aperture.  This sky level is then simply scaled by the relative glint/direct-sunlight ratio to provide the estimated scattered sunlight contribution to the total sky levels measured.
 
 For example, the n3c61f field has the smallest SEA, and incurs the strongest scattered sunlight effects.  From the ray-tracing tests appropriate to the spacecraft attitude used to observe this field, we observe a glint that has at most $1.8\times10^{-3}$ of the solar flux and a $9^\circ$ elevation as observed from the LORRI aperture\footnote{In examining detailed photographs of the LORRI aperture latch mechanism, we think it likely that the model is also incorrectly including a scattered-light contribution from it.}.  At ${\rm SEA=81^\circ},$ direct sunlight produces a 21 DN background in 30s.  Multiplying  this by the relative flux of the glint, we estimate that the glint produces a scattered sunlight background of 0.04 DN in 30s.  This is $\sim20\%$ of the final dCOB signal that we will recover by the analysis to be described later.  The ZL138c and ZL138d fields have over an order of magnitude smaller scattered sunlight components, and the remaining four fields have no detectable scattered  sunlight entering the camera at upper limits two orders of magnitude  smaller than that in the n3c61f field. Because only one field has a scattered sunlight component at even relatively small significance, we have chosen to not to correct the average sky for this effect.
 
 \subsubsection{Glittering Ammonia Crystals?}
 
 Fine guidance control during long LORRI exposures is done by frequent firing of the New Horizons attitude control thrusters. Thrust is provided by catalytic decomposition of hydrazine (${\rm N_2H_2}$), which generates a hot gas plume comprising ${\rm N_2},$ ${\rm H_2},$ ammonia (${\rm NH_3}$), and a small fraction of water, which may be present in the hydrazine at the trace level.  Four thrusters eject plumes parallel to the LORRI optical axis, raising the concern that as the gas cools, ammonia-ice crystals might form that would then scatter sunlight into LORRI.
 
 Full exploration of this problem is beyond the scope of this paper, but a quick examination of the thruster parameters suggests that by the time the plume has expanded and cooled enough to allow ice crystals to form, the molecular free mean path is too large to allow the kinetic molecular interactions that would form and grow crystals. The amount of fuel consumed during an exposure is small in any case, and it appears unlikely that an exhaust cloud with an optical depth sufficient to affect the LORRI sky level would form.
 
 \subsection{LORRI Initial Background Fade}

\begin{figure}[htbp]
\centering
\includegraphics[keepaspectratio,width=3.5 in]{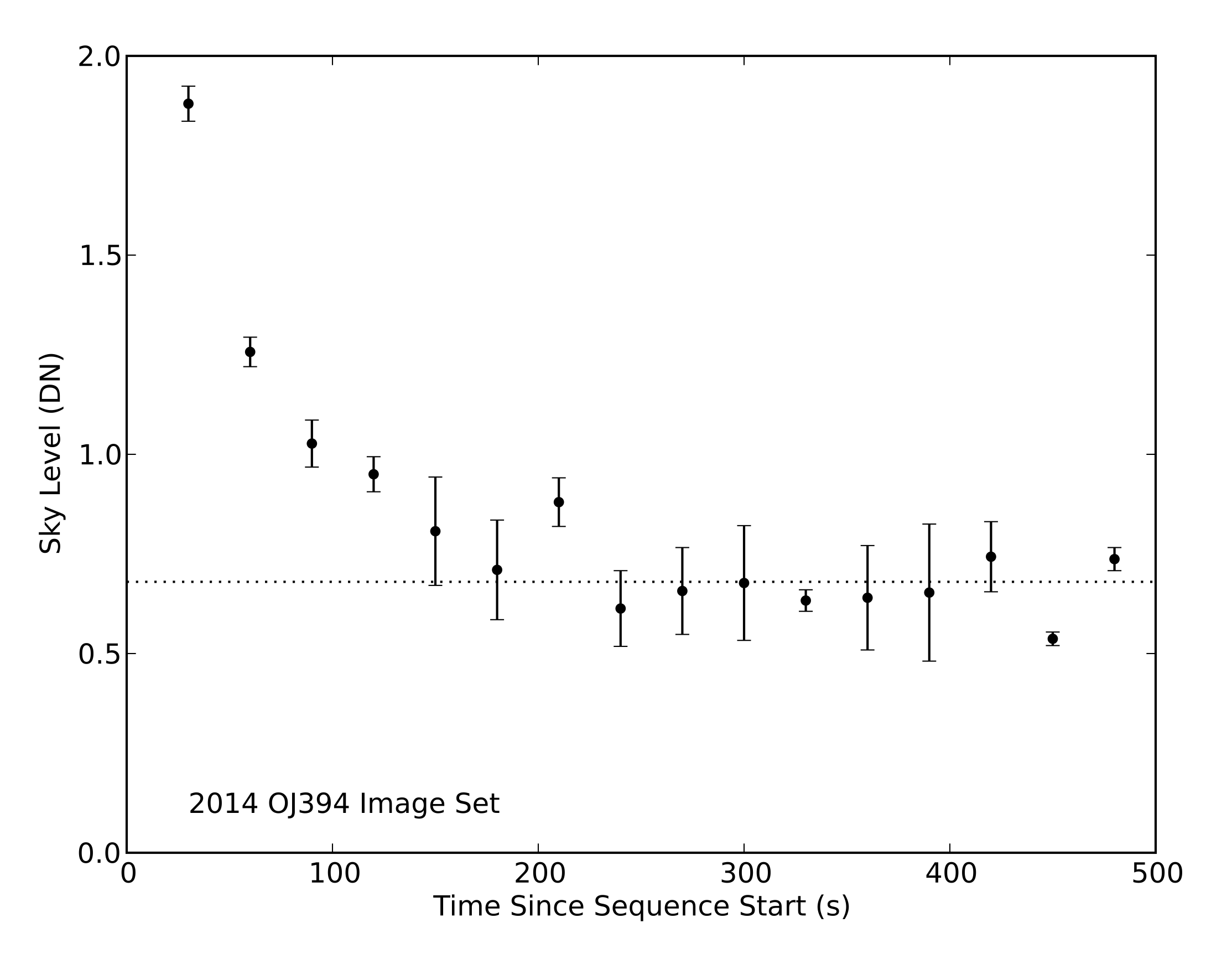}
\caption{The average reduced sky level is shown as a function of time based on three sequences of 16 consecutive 30s LORRI images obtained shortly after the camera was powered up, obtained from a program to observe KBO 2014 OJ394. The first four images in all sequences clearly have a higher background level than the subsequent images do. The dotted line shows the average sky level determined by excluding the first five images in the sequences. Error bars are the error in the mean for the three exposures contributing to each point ($\pm1$-$\sigma$ error bars are shown in this and all other figures).}
\label{fig:background}
\end{figure}

The need for a long run of images in a data set is to counter a recently discovered background effect within the camera \citep{lorri2}. The effect is manifested as elevated bias and sky levels, which are most pronounced in the first exposure in a sequence, but require a $\sim150$s interval (Figure \ref{fig:background}) to decay fully to a constant level. This effect appears to be associated with the power-on activation of the camera, which typically occurs 30s before the start of an imaging sequence. Notably, it is not seen at the start of sequences that occur well after the initial sequence, but still within the same operational interval over which LORRI was continuously powered. With sequences of 30s exposures, we chose to discard the first five exposures following the activation of the camera to avoid the effects of this phenomenon.

\subsection{The Sky Fields}

\begin{figure*}[htbp]
\centering
\includegraphics[keepaspectratio,width=6 in]{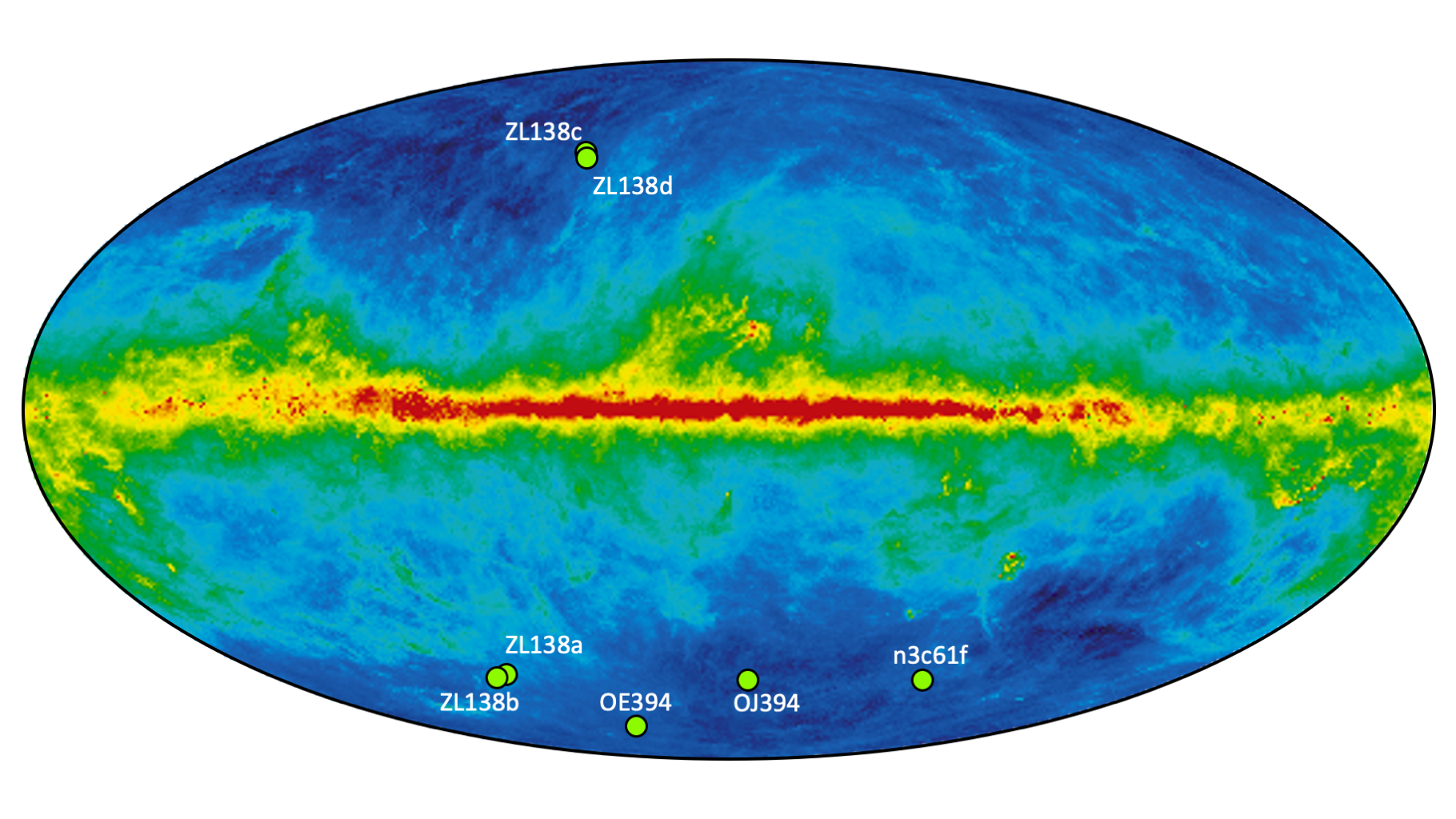}
\caption{The location of the seven LORRI fields used in this work, shown projected on the extinction map derived by \cite{sfd} from the IRAS 100$~\mu$m all-sky survey. Note that  the symbols for the ZL138c and ZL139d fields nearly overlap.}
\label{fig:nhskymap}
\end{figure*}

We identified seven LORRI data sets that satisfied all the constraints. The sequence parameters are given in Table \ref{tab:obs}. The field locations with respect to a galactic extinction map derived from the IRAS 100$~\mu$m all-sky survey \citep{sfd} are shown in Figure~\ref{fig:nhskymap}. Three of the  seven data sets were provided by DKBO observations. The DKBO data sets comprise several imaging sequences of three KBOs, each taken over an interval of 14 to 48 hours.  The LORRI pointing changes over each set as needed to track the KBO, but varies by only about half a LORRI field. The background $100~\mu{\rm m}$ flux does not vary significantly over each set as the pointing changes (as will be discussed in detail in the analysis section). Each set can thus be combined into a single measurement.

The remaining four data sets are single-pointing image sets generated by a zodiacal light program designed to provide a quick test for obvious dust within the plane of the cold classical Kuiper Belt. The program comprised four fields at the same high galactic latitude ($|b|\sim60^\circ$), but with two fields at ecliptic latitude $\beta\sim0^\circ$ and two at $\beta\sim50^\circ.$ The target fields were also chosen have low surface densities of bright stars. A sequence of eight consecutive 30s LORRI exposures were obtained of each field.  The reduction of these sequences led to the discovery of the initially high instrument background shown in Figure \ref{fig:background}.  

\begin{deluxetable*}{lrrrrrccccc}
\tabletypesize{\scriptsize}
\tablecolumns{11}
\tablewidth{0pt}
\tablecaption{Sky Fields and Imaging Sequences}
\tablehead{\colhead{}&\colhead{}&\colhead{}&\colhead{Solar}
&\colhead{}&\colhead{}&\colhead{}&\colhead{$I_{100}$}&\colhead{Date}&\colhead{$r$}
&\colhead{}\\
\colhead{Program}&\colhead{RA}&\colhead{Dec}&\colhead{Elong.}
&\colhead{$b$}&\colhead{$\beta$}&\colhead{$E(B-V)$}&\colhead{${\rm MJy~ sr^{-1}}$}&\colhead{(UT)}&\colhead{(AU)}
&\colhead{$N_{seq}/N_{im}$}}
\startdata
OE394  &$  1.7790$&$-17.7780$&$110.1$&$-76.1$&$-17.0$&0.029&1.02&2018-08-20&42.2&6/78\\
OJ394  &$348.0611$&$-41.6360$&$125.0$&$-65.1$&$-33.2$&0.010&0.53&2018-08-22&42.2&9/120\\
n3c61f &$ 33.4069$&$-50.7529$&$ 96.3$&$-61.8$&$-58.1$&0.017&0.89&2018-09-01&45.2&15/90\\
ZL138a &$358.2428$&$ -0.5183$&$108.3$&$-59.8$&$  0.2$&0.025&0.73&2019-09-04&45.3&1/8\\
ZL138b &$  0.8066$&$  0.2915$&$105.6$&$-60.2$&$ -0.1$&0.031&0.97&2019-09-04&45.3&1/8\\
ZL138c &$224.9875$&$ 36.2331$&$ 98.1$&$ 61.4$&$ 50.3$&0.012&0.68&2019-09-04&45.3&1/8\\
ZL138d &$226.4865$&$ 35.2979$&$ 99.6$&$ 60.3$&$ 50.0$&0.011&0.70&2019-09-04&45.3&1/8\\
\enddata
\tablecomments{Column (1) partial program name (`ZL' stands for ``ZodiacLight"), (2) Sky field RA (2000), (3) Dec (2000), (4) solar elongation, (5) galactic latitude, (6) ecliptic latitude, (7) Reddening from \citet{sf11}, (8) Average 100$~\mu$m flux for the field from the \citet{iris} maps, with correction for residual zodiacal light, and the 0.78 ${\rm MJy~ sr^{-1}}$ cosmic IR background \citep{cib1, cib2} subtracted (see $\S\ref{sec:dgl}$), (9) UT starting date, (10) New Horizons distance from the Sun, (11) number of sequences / total number of images. All angles are in degrees. For the first three fields, the coordinates given are an average over all sequences. The E(B-V) and 100$~\mu$m flux values were obtained from the IRSA archive dust tool at https://irsa.ipac.caltech.edu/applications/DUST/.}
\label{tab:obs}
\end{deluxetable*}

\section{Measuring the Sky With LORRI}\label{sec:measure}

\subsection{Image Reduction}

The sky levels are less than 1 DN in the nominal 30s exposures.  The reduction of the images thus requires attention to a number of subtle effects that are only important at this level.  Rather than using calibrated (``Level 2'') images produced by the standard LORRI pipeline operated by the New Horizons project, we designed a custom reduction of the raw (``Level 1") images to optimize accurate recovery of the faint sky signal.

\subsubsection{Bias Level Determination}

The LORRI detector, like all CCD cameras, records a signal superimposed on a background voltage or bias level generated by the camera electronics. The bias level varies slightly over a sequence of exposures, thus it must be determined for each image.  In LORRI $4\times4$ mode the bias is measured from a single column of un-illuminated pixels, which comprises an extra pixel in each row that follows the 256 illuminated pixels.  In the standard LORRI pipeline the overall bias level for the image is taken as the median DN level of the bias column.  This forces the bias to an integral value, however, introducing a potential error as large as 0.5 DN in its determination. We instead measured the bias by fitting a gaussian to the peak of the histogram of DN values in the bias column, providing a measurement accurate to a fraction of a DN. 

\subsubsection{Measurement of the LORRI Dark Current}

The bias column also integrates any dark current present in the detector over the duration of the exposure. As such, any constant dark-level over the field will be treated as part of the bias level and removed with it. As is discussed in \citet{lorri2}, the dark current in a LORRI $1\times1$-mode pixel is estimated to be $4\times10^{-3} e^-~{\rm s^{-1} pixel^{-1}}$ or $\sim0.1$ DN/pixel in a 30s exposure in $4\times4$ mode, based on the manufacturer's specifications and the CCD operating temperature. A modest contribution to the background at this level should be well characterized by the bias column.

Higher dark levels, however, place higher demands on the accuracy of the underlying assumption that the bias column truly witnesses the average dark current appropriate for the active imaging area of the LORRI CCD.  Unfortunately, without a dark shutter we cannot obtain true dark calibration exposures over the full field in flight; however, we have been able to obtain calibration data that demonstrates that the dark level measured by the unilluminated bias column is indeed close to the expected level.

In the ZeroDark65 LORRI sequence, which operated on the spacecraft in June 2020, we obtained 32 pairs of 65s blank-sky\footnote{The astronomical field was selected to have a low density of stars, as well as avoiding any bright stars that might become over-exposed in 65s. To minimize spacecraft and down-link resources, however, the  exposures were not done with fine-pointing control and only the bias columns were down-linked.   This sequence thus provides no useful  sky observations.} exposures immediately followed by a 0s image  obtained 1s later. While the  electronic bias level can drift over a long imaging sequence, we have found it to be stable by a fraction of a DN between consecutive exposures. If the electronic bias level were truly constant, the difference between the measured bias plus dark levels recorded by the unilluminated columns in the pair of images would provide a direct measure of the dark level.  With 32 such pairs the measurement noise and any random bias variations should average out.  Indeed, the bias level drifted systematically by no more than 0.5 DN over the full duration of the ZeroDark65 sequence.

\begin{figure}[htbp]
\centering
\includegraphics[keepaspectratio,width=3.5 in]{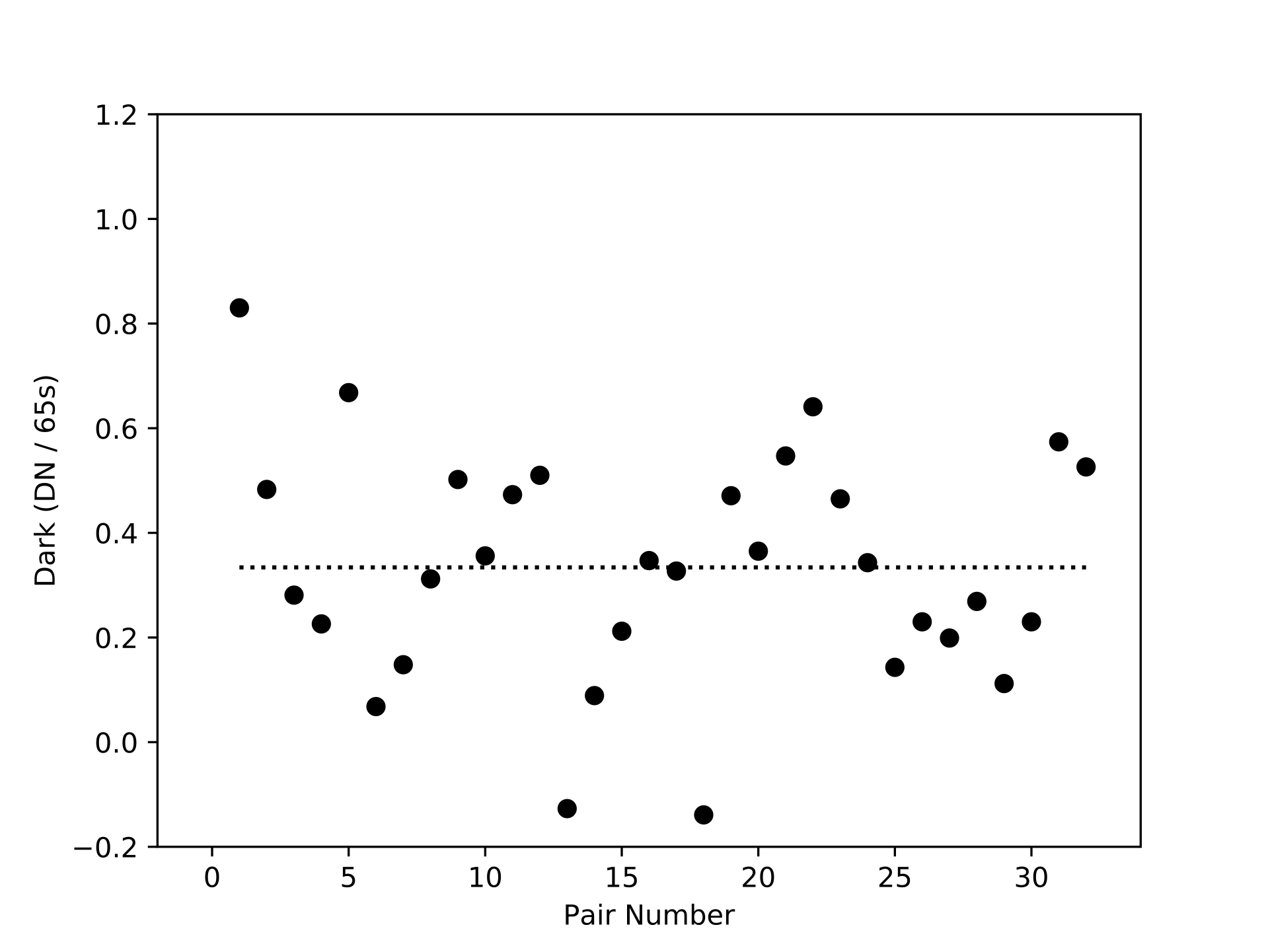}
\caption{The 32 dark current measurements obtained in the ZeroDark65 calibration sequence are plotted in units of DN in 65s.  The dashed line gives the average dark current of $0.334\pm0.039$ DN in 65s.}
\label{fig:zerodark}
\end{figure}

Figure \ref{fig:zerodark} shows the dark current measures derived by differencing the 32 65s and 0s image pairs. The average level measured is $0.334\pm0.039$ DN in 65s, or $0.154\pm0.018$ DN in 30s.  Referenced to the LORRI $1\times1$ mode, this is $6.2\times10^{-3} e^-~{\rm s^{-1} pixel^{-1}},$  which is within the range of the CCD manufacturer's specified performance.  This modest dark level should be completely corrected by the measured bias level subtraction described in the previous section.

\subsubsection{Correction for Bias structure}\label{sec:bias}

A variety of coherent and repeatable patterns are evident at low signal levels in LORRI images.  The standard LORRI pipeline subtracts a ``super-bias'' image constructed from an average of a large set of bias frames obtained shortly after the New Horizons spacecraft was launched but before the LORRI aperture was opened. We instead used a revised ``super-bias'' image based on extremely short in-flight images obtained with the camera, processed to correct only large-scale features in the bias frame.

\begin{figure}[htbp]
\centering
\includegraphics[keepaspectratio,width=3.5 in]{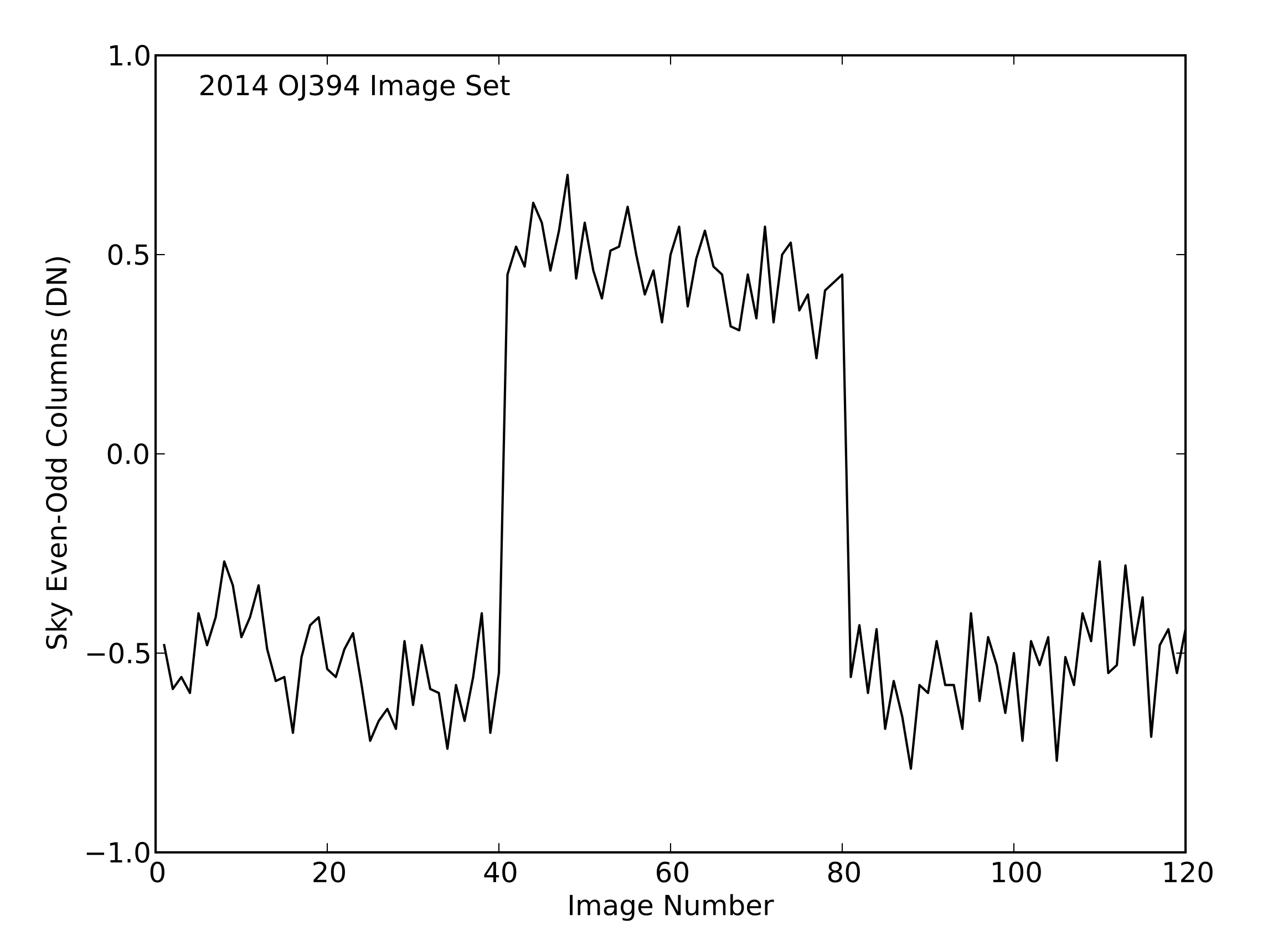}
\caption{The difference the sky levels measured from the even CCD columns as compared to the odd columns is shown as a function of image number counting from the start of the 120 observations of the KBO 2014 OJ394. The bias-reference column is an odd-numbered column.  The 0.5 DN offset of the even columns makes a ``jail bar" background pattern in the LORRI images.  The sign of the offset appears to be set randomly when the camera is powered up.  The set of 120 OJ394 images was divided into three sets of 40 images obtained in different ``visits." Note that the sign of the offset in the second visit is opposite that in the first and last visits.}
\label{fig:jail}
\end{figure}

\begin{figure}[htbp]
\centering
\includegraphics[keepaspectratio,width=3.5 in]{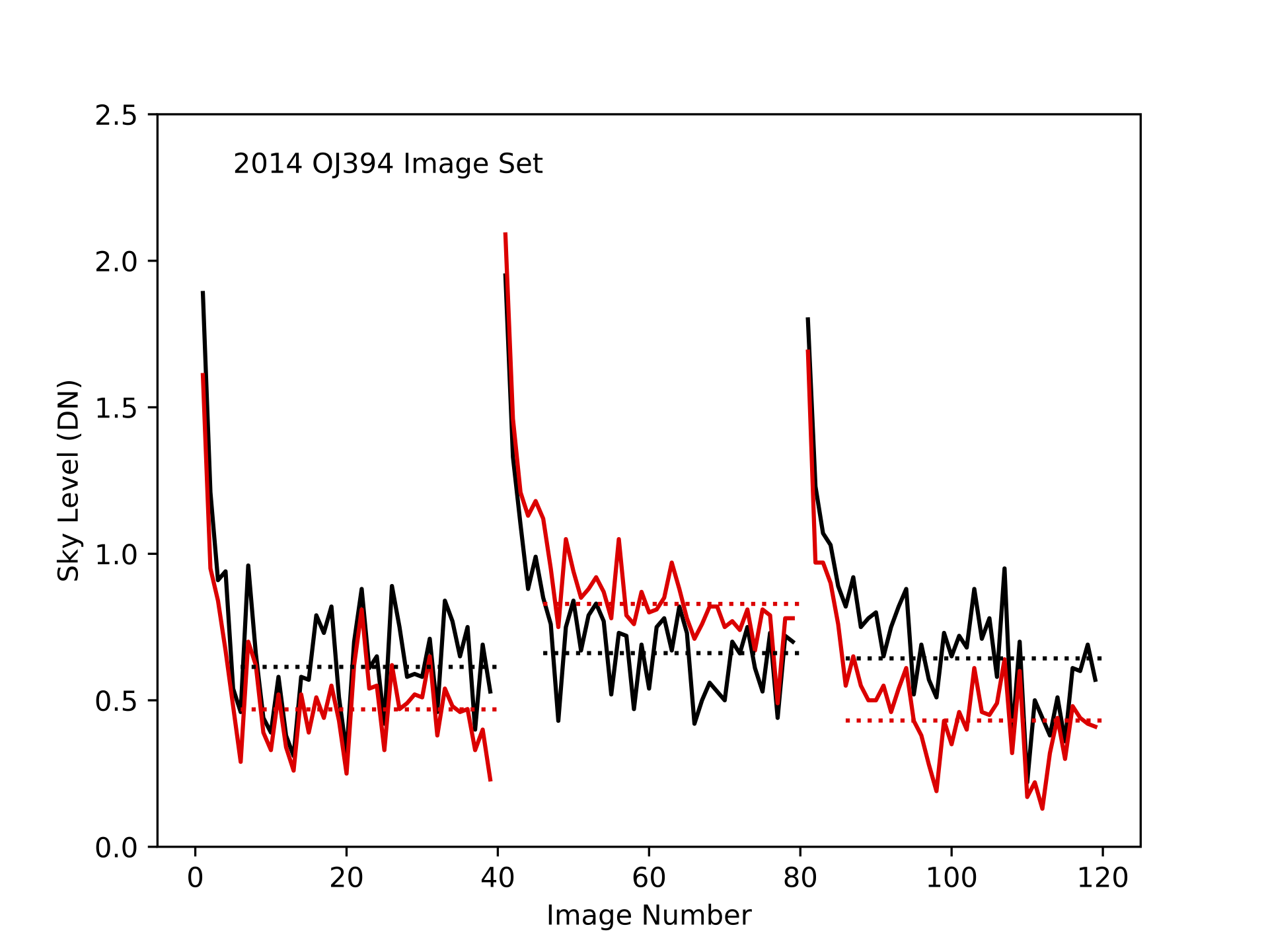}
\caption{The black/red traces shows the sky level as a function of image-number with/without the jail-bar correction applied for the 2014 OJ394 observations, which were done in three visits, each comprising 40 images. The image number reflects the temporal order of the images; however, the three visits were separated by several hours during which the LORRI camera was powered off. The dashed lines show the average sky level for each of the three visits As shown in Figure \ref{fig:jail}, the second of three visits has a jail-bar pattern of opposite of that for the flanking visits. Note the excellent agreement of the average levels once the correction as been applied.  The spike in the sky level at the start of each visit is the same phenomenon shown in detail in Figure \ref{fig:background}.}
\label{fig:jail_error}
\end{figure}

LORRI frames exhibit a low-level ``jail bar" pattern in which the average level of the even versus odd columns differ by 0.5 DN \citep{lorri2}.  This pattern is not treated in the standard pipeline and can strongly affect determination of the sky level. Since the bias column has odd parity (it is read out
as column 257, following the 256 illuminated columns), the bias level determined from it is strictly valid only for the odd columns. A key aspect of the pattern is that the even columns vary from sequence to sequence in being either $-0.5$ or $+0.5$ DN offset from the odd columns.  The sign of the offset appears to stay constant over a given LORRI power-on interval, but changes randomly from different operational cycles (Figure \ref{fig:jail}). Without correcting for this effect, the measured sky level will vary between being $-0.25$ DN too low or $+0.25$ DN too high. This effect is evident as systematic differences of 0.5 DN between different sequences obtained of the same field, as is shown in Figure \ref{fig:jail_error}.  An error of this size is close to the full value of the sky signal, itself.

The solution to the jail-bar offset is simple. Given the faint background levels in the present image sets, it is easy to determine the sign of the 0.5 DN offset of the even from the odd columns, and thus apply the appropriate correction. Figure \ref{fig:jail_error} shows the same data set before and after the jail-bar correction was applied.  The agreement of the average sky level between visits with opposite jail-bar parities is now excellent.

Lastly, the bias level appears to show random low-level variations over the duration of the image readout, which is evident as horizontal streaking in the image backgrounds. The bias column forces this pattern to have zero-mean at the high column-number margin of the images, but the bias may also exhibit a slow low-amplitude drift along the CCD rows, such that the overall mean of the bias background alone may have a small non-zero value.  At present this pattern is not corrected, and it may contribute to some of the random exposure-to-exposure variations in the sky level, such as those visible in Figure \ref{fig:jail_error}.

To validate these reduction steps, we tested them on two 0s LORRI $4\times4$ exposures obtained on July 13, 2019 for routine performance monitoring of the camera.  Using the histogram-estimation methodology to be described in $\S\ref{sec:histog},$ we measured the average signal level as $0.02\pm0.06,$ DN, demonstrating that we measure  zero signal in data expected to have a sky level of zero.

\vskip 30pt
\subsubsection{Charge-Smearing Correction and Flat-Fielding}

After the average bias level is subtracted from the image, and the bias structure corrections described above are completed, a ``charge-smearing" correction is applied \citep{lorri2}.  In brief, because LORRI does not have a camera shutter, light from an astronomical source will still deposit charge in the LORRI CCD as it is being read out, and as the CCD is clocked in advance of the exposure to erase or scrub-out charge deposited prior to the start of the exposure.  The exposure is made by simply stopping the charge clocking for its duration. The effect of charge being deposited both before and after the nominal exposure interval is to generate charge trails in the image rows above and below a bright source.  The distribution of smeared charge  is corrected prior to flat-fielding.

The amplitude of the smeared charge in any given pixel in the
same column as a bright source can be estimated from the flux of the source, scaled to the time required to clock out a single row of the image.  For the LORRI $4\times4$ mode this is only 0.047 ms.  In the present case, for a bright star that just saturates at 4095 DN in 30s, the smeared charge is only 0.005 DN/pixel, which is only $\sim1\%$ of the typical sky levels that we are concerned with. Even so, we still apply the smear correction using the new algorithm discussed in \citet{lorri2}.  We note that we flag the cosmic-ray hits in any given image in advance of the smear correction, as these are instantaneous sources that will not be smeared.  Including the CR hits in the correction will induce a small negative bias to other pixels in the affected column.

\begin{figure*}[htbp]
\centering
\includegraphics[keepaspectratio,width=5.5 in]{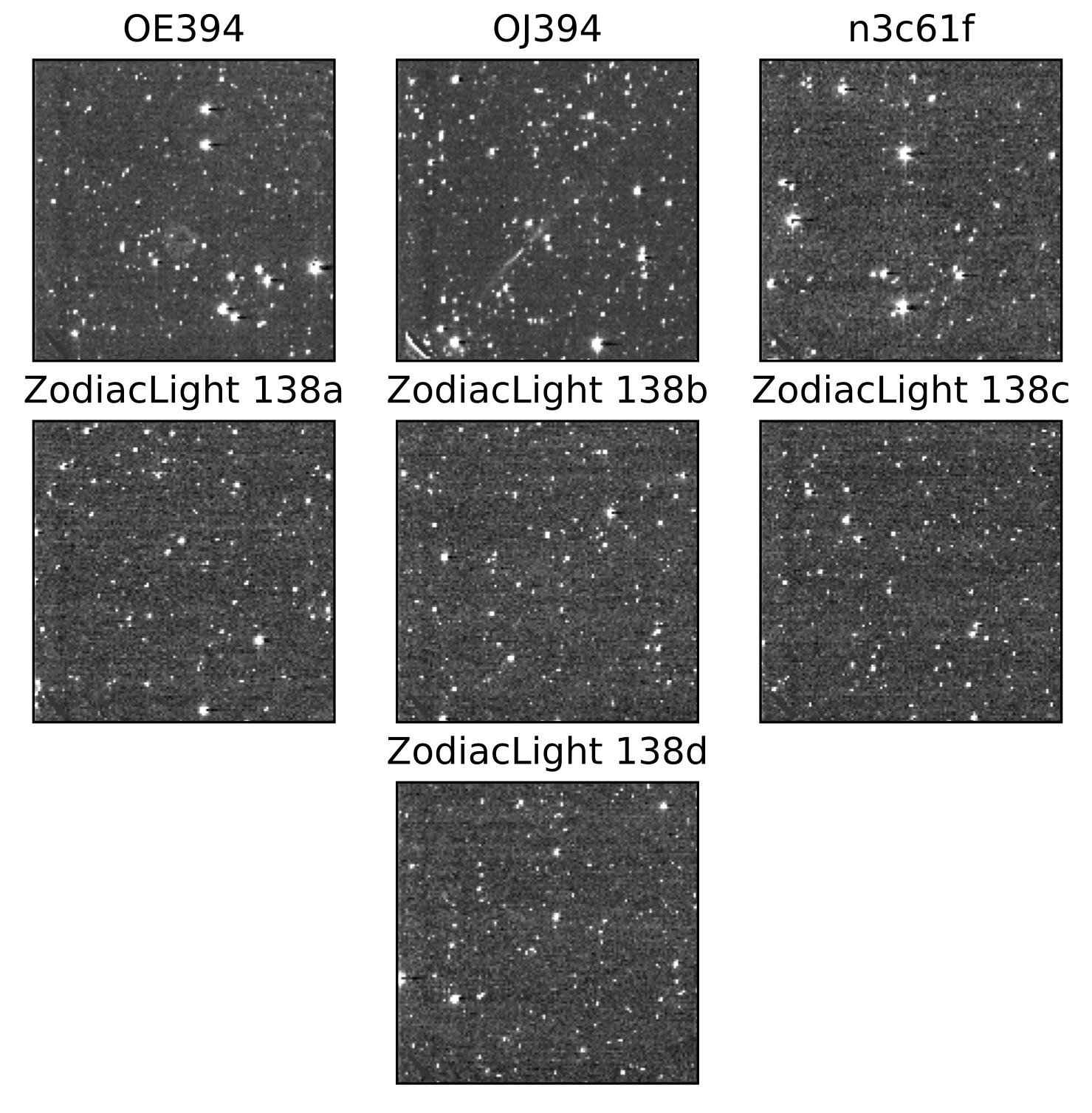}
\caption{Image stacks of the seven fields are shown.  The stretch is 10 DN, starting at $-2$ DN. North is at the top and east to the right (i.e., x-axis is inverted from the conventional orientation). Multiple stacks at slightly different pointings are available for the top three fields; a representative stack at one position is shown. Scattered light ``ghosts'' are evident in the OE394 and OJ394 fields, and pixels affected by them were excluded from the analysis.}
\label{fig:sky_image}
\end{figure*}

The last image reduction step is the flat-fielding sensitivity correction, which is done with the standard pipeline procedures and products.  This is a minor correction.  The response of the CCD is highly uniform over its area, with rms sensitivity variations of only 0.9\%. The final reduced images are shown in Figure \ref{fig:sky_image}.

\subsection{Measuring the Sky Level}

\subsubsection{Exclusion of Non-Sky Flux Sources} \label{sec:mask}

For the sparse high galactic-latitude fields in the present sample, only a small minority of the pixels in a given image are encompassed by detectable astronomical sources.  The sky level can easily be derived from the peak of the histogram of pixel-intensities, once care is taken to account for the low-intensity wings of stars, galaxies, cosmic rays, hot-pixels, optical ghosts, and any other objects present in the field.  In practice, we find that derivation of the peak-intensity is not sensitive to how the pixels containing sources and CCD defects are masked nor to the algorithm used to estimate the location of the peak.

Since we are concerned only with the distribution of pixels within a few DN above or below the nominal zero-level, the main reduction step is to flag and exclude areas of the image encompassed by objects, rather than trying to subtract or correct for them.  As we had several images to work with for each field, cosmic rays could be recognized as statistical outliers over a given sequence of images.  A list of hot pixels down to the 2 DN level were recognized from pixels that showed consistent positive offsets above the background over the 120 images in the 2014 OJ394 set.  The astronomical pointing varied markedly over the various visits and sequences, such that over the full data-set, all pixels sample the background the majority of the time. Hot pixels were excluded from the intensity histogram generated for each image, regardless of their strength in any given image, as were all pixels recognized as cosmic-ray events.

Two approaches with differing levels of aggressiveness were used to recognize and exclude detectable astronomical objects from the intensity histogram produced for any image. As noted below, however, both produced only modest corrections to the sky level derived without any processing of the objects.

Our adopted strategy was to flag all pixels of intensity 8 DN or more above the zero-level established by the bias, as well as all neighboring pixels within 12 arcsec from them. For point sources, this corresponds to a photometric limit of $m_v=19.1,$ which we detect at 5-$\sigma$ significance in the 30s exposures. Despite the high-level  of object rejection achieved, however, the average decrease in the sky before and after the masking procedure was only 0.03 DN in a 30s exposure; in our sparse fields, the histogram-based sky algorithm (to be described in the next section) is largely invulnerable to bright point sources even without masking them out.  At the same time, however, while the 12 arcsec exclusion radius eliminates the visible wings of even the brightest objects within the image, their faint extended wings outside this radius will still contribute to the overall image background. As will be discussed in detail in $\S\ref{sec:ssl},$ stars in the field with $m_v<19.1$ are included in the integrated scattered starlight correction.

As a check on this procedure, we also generated a mask for each sequence by co-adding all the frames in the sequence, and then lightly smoothing the stack to recognize objects considerably closer to the detection limit of the camera.  This procedure greatly increased the set of pixels excluded from the histogram in any image, but effected only an additional average decrease of 0.04 DN in the sky level.  Because the number of images that could be stacked for any field was highly heterogeneous over the entire sample, we chose to use the less aggressive algorithm, which could be applied uniformly to the individual 30s exposures.

An important caveat is that with the modest aperture of LORRI
and the modest 30s exposure time, the present images are extremely shallow compared to those provided by deep imaging surveys conducted with the Hubble Space Telescope and large ground-based telescopes. An important part of the interpretation of the present sky levels, which we will take up in a later section, is to estimate the integrated contribution of galaxies and stars fainter than the $m_v=19.1$ detection limit for any single object.

Lastly, we noted that some fields were affected by faint ghosts of bright stars well outside the LORRI field  of view.  Again, pixels affected by the ghosts were excluded from the analysis.  We also excluded the first 32 columns of the LORRI CCD for which the super-bias correction appeared to slightly depress the background compared to rest of the field.

\subsubsection{Determining the Peak Location of the Intensity Histogram}\label{sec:histog}

\begin{figure}[htbp]
\centering
\includegraphics[keepaspectratio,width=3.5 in]{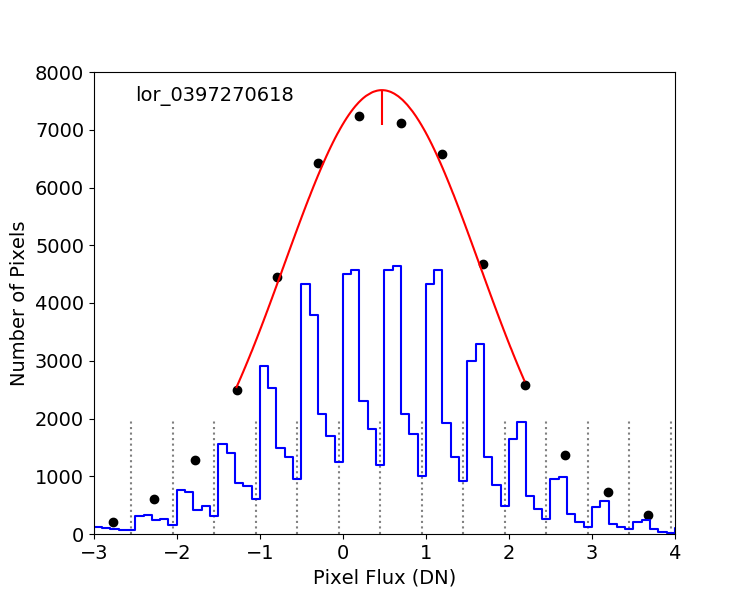}
\caption{Measurement of the sky level from the histogram of pixel intensities is demonstrated for LORRI image  lor\_0392770618, one the images drawn from the 2014 OJ394 dataset. The blue trace is the histogram of the image at 0.1 DN resolution (scaled up by $2\times$ for clarity). The histogram has peaks spaced by 0.5 DN due to the ``jail-bar" correction. This histogram is binned at 0.5 DN resolution to generate the black solid points.  The bins (separated by gray dotted lines) are offset to reflect the fractional-DN value of the bias level. The flux positions of the black points are calculated by an intensity centroid of the pixels in a given bin.  The red line is a gaussian fitted to the black points as shown.  The sky level is given by the location of the peak of the gaussian.}
\label{fig:histogram}
\end{figure}

The sky level for any given image is determined from the peak of its pixel-intensity histogram.  The  histograms were generated in a way that best preserved the information content of the images (Figure \ref{fig:histogram}). For pixels sampling the low sky levels most of the image reduction steps only slightly altered the raw integral DN values  generated by the camera, thus care had to be  taken in how the signal was binned to generate the histogram.  The most important consideration was to preserve the 0.5 DN offset correction applied to correct the jail-bar pattern, and the fractional DN portion of the bias level measurement. The procedure adopted was to use 0.5 DN-wide bins for the histogram, and to phase the centers of the bins to respect the fractional part of the bias-level subtracted from the initially integer DN values. Lastly, the average intensity of all the pixels within a given bin was used to define the center of the bin, rather than the simple mid-point of the bin interval.

We explored three algorithms for measuring the precise intensity of the histogram peak. All approaches used the bin with the maximum occupation number and bins on either side of it that had occupation numbers falling just below half of the peak value.   The first procedure was to fit a parabola to the occupation numbers a function of bin average-intensity.  The second was to fit a parabola to the logarithm of the occupation numbers, which is equivalent to fitting a gaussian to the bins. Lastly, a simple centroid of the peak bins was calculated. All three measures agreed at the 0.01 DN level. We adopted the gaussian-fit measures as they produced the smallest scatter in measurements over a large sample.  

\begin{deluxetable}{lrccc}
\tabletypesize{\scriptsize}
\tablecolumns{5}
\tablewidth{0pt}
\tablecaption{Measured Total Sky Levels}
\tablehead{\colhead{Program}&\colhead{N}&\colhead{DN}&\colhead{$e^-$}&\colhead{${\rm ~nW~m^{-2}~sr^{-1}} $}}
\startdata
OE394\hskip 80pt  & 63 &$0.758\pm0.025$&$14.7\pm0.49$ &$37.52\pm1.24$\\      
OJ394\hskip 80pt  &105 &$0.613\pm0.012$&$11.9\pm0.23$ &$30.34\pm0.59$\\      
n3c61f\hskip 80pt & 15 &$0.803\pm0.034$&$15.6\pm0.66$ &$39.75\pm1.68$\\      
ZL138a\hskip 80pt &  3 &$0.737\pm0.053$&$14.3\pm1.03$ &$36.48\pm2.62$\\     
ZL138b\hskip 80pt &  3 &$0.804\pm0.044$&$15.6\pm0.85$ &$39.80\pm2.18$\\     
ZL138c\hskip 80pt &  3 &$0.641\pm0.045$&$12.4\pm0.87$ &$31.73\pm2.23$\\     
ZL138d\hskip 80pt &  3 &$0.721\pm0.028$&$14.0\pm0.54$ &$35.69\pm1.39$\\
\enddata
\tablecomments{Column (1) partial program name, (2) number of images used, (3) average sky level DN/pixel in 30s, (4) sky in electrons/pixel in 30s, (5) sky in intensity units.}
\label{tab:skylevs}
\end{deluxetable}

The sky levels for each field are given in Table \ref{tab:skylevs}. They are averages of the individual levels of all images available for a given field, excluding the first five images obtained following the powering-up of LORRI.  The errors are the statistical errors of the mean.  The typical error in any single image is 0.15 DN or $7.5{\rm nW~m^{-2}~sr^{-1}},$ but does appear to vary somewhat between sequences.  Conversion to flux/solid-angle values is provided by equation (8) in \citet{lorri2}, assuming the RSOLAR zeropoint defined in that paper. Direct conversion of a sky level, $S,$ in DN / (LORRI $4\times4$ pixel) in 30s, assuming a pivot wavelength of 0.608 $\mu$m, is
\begin{equation}
    \lambda I_\lambda = 49.5~S~{\rm nW~m^{-2}~sr^{-1}}.
\label{eq:s2iconv}
\end{equation}

\vskip 14pt
\section{Decomposing the Sky}\label{sec:decomp}

The sky signal as measured reflects an integral over a number of possible contributions:
\begin{itemize}
\item integrated light from faint stars and galaxies below the point-source detection-limit in the LORRI images,
\item scattered light from bright stars and galaxies in and around the LORRI field,
\item diffuse Milky Way starlight scattered by infrared ``cirrus,"
\item scattered sunlight from dust within or beyond the Kuiper Belt,
\item a diffuse cosmic optical background (dCOB) not associated with any extragalactic source population presently known, and
\item unaccounted for dark current or scattered light in the LORRI camera.
\end{itemize}
The goal in this section is to remove known sources from the integrated sky, constraining the contributions of unknown components. Of the components listed above, it is straight forward to estimate the integrated contributions of faint stars or galaxies to the LORRI sky. The diffuse galaxy light (DGL) contributed by the Milky Way starlight is more problematic, but can be estimated using the ${\rm 100~\mu m}$ thermal emission from infrared ``cirrus" dust features to predict the optical-band light scattered by them. Significant sky-signal remaining after these components have been removed represents ``unknown" and possibly novel sources, or an unknown artifact produced by the camera or spacecraft. 

At the outset, we are struck by the fact that the seven total sky measures are  highly uniform. The mean sky level is $33.2\pm0.5{\rm~nW~m^{-2}sr^{-1}},$ with dispersion $3.7{\rm~nW~m^{-2}sr^{-1}}$ or only 11\% of the mean.  While the fields are all at high galactic latitude, they do cover a wide range of ecliptic latitude --- there appears to be little room for any scattering component associated with the plane of the ecliptic, as we discuss quantitatively at the end of the  next section.  Lastly, as we will discuss in $\S\ref{sec:comparison},$ this total sky level falls markedly below a number of {\it final} COB measures presented in the literature even before the corrections discussed in the following sections are applied.  In magnitude units, the average total sky corresponds to 26.0 V mag arcsec$^{-2}$, or more than $10\times$ fainter than the darkest sky available to the {\it Hubble Space Telescope}. 

\begin{figure*}[htbp]
\centering
\includegraphics[keepaspectratio,width=6.0 in]{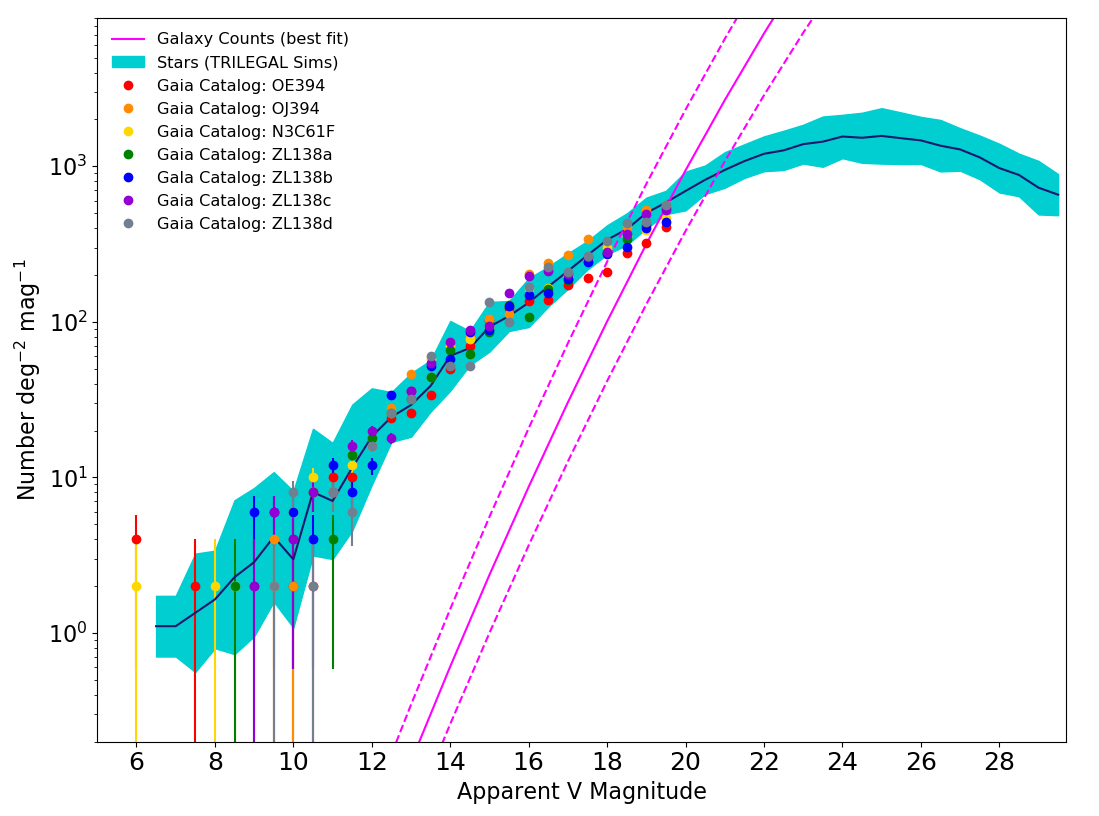}
\caption{The number of stars per square degree per magnitude in the V-band. The results for the {\tt TRILEGAL} \citep{tril2005, trilv16} simulations are represented by the blue shaded region. The vertical width of the {\tt TRILEGAL} curve represents the variation in the star number counts in the seven different fields. The colored data points are the star counts in the {\it Gaia} DR2 catalog \citep{gaia2016,gaia2018} over the range $5 \le {\rm G} < 19.5$ from the 1 square degree regions around each LORRI pointing. For reference, we show the galaxy counts (magenta curve) based on a number of extragalactic sky surveys and deeply imaged sky fields (see \S~\ref{ssec:fgal}). Apparent magnitudes are on the AB system.}
\label{fig:vstars}
\end{figure*}

\begin{figure*}[htbp]
\centering
\includegraphics[keepaspectratio,width=6.2 in]{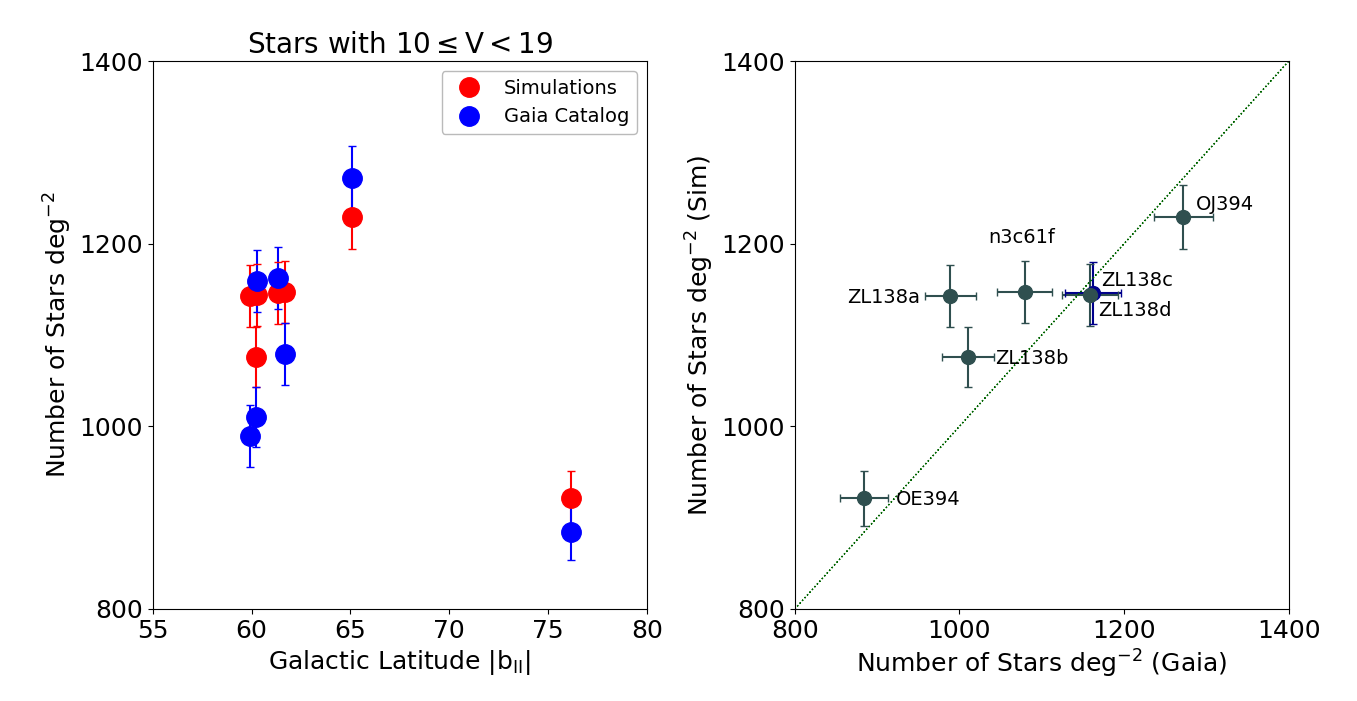}
\caption{The field-to-field variation in number of stars per square degree over the interval $10 \le {\rm V} < 19$ is shown for the seven fields for both the {\it Gaia} DR2 catalog and the simulated star fields generated by the {\tt TRILEGAL} package. The left hand plot shows the star count density as a function of galactic latitude. The right hand plot shows the direct comparison between the number of stars observed and predicted, respectively, by {\it Gaia} and the {\tt TRILEGAL} model.}
\label{fig:svar}
\end{figure*}

\subsection{Integrated Light from Undetected Faint Stars}\label{ssec:fstar}

The background light from stars within the fields below the detection limit of the LORRI images is estimated using version 1.6 of the {\tt TRILEGAL} Milky Way model \citep{tril2005, trilv16}. For each LORRI field, we generate a one square-degree simulation centered on the boresight coordinates given in Table~\ref{tab:obs}. The simulations are performed using a limiting magnitude of J = 30 and the default parameters for the four main Galactic components: thin disk, thick disk, halo, and bulge. Given the high galactic latitudes of the observations used here, the simulations only include stars from the disk and the halo. As the LORRI field covers $0\adeg29\times 0\adeg29$, running a one square-degree simulation effectively generates $\sim12$ images worth of stars for each pointing, which gives us ample mitigation against sample variance. The variation seen in the simulated star counts when running {\tt TRILEGAL} for a given sky field multiple times using identical model parameters is less than 0.7\%.

We derive the number counts of stars as a function of magnitude for the UBVRIZJ passbands to encompass the LORRI sensitivity range. We apply the following (Vega - AB) magnitude conversions, as needed: $-0.7194$ (U), $+0.123$ (B), $-0.017$ (V), $-0.211$ (R), $-0.450$ (I), $-0.573$ (Z), and $-0.940$ (J). The star number counts in the V-band is shown in Figure~\ref{fig:vstars}. We also show in this figure the actual star counts from the DR2 {\it Gaia} catalog \citep{gaia2016,gaia2018} in the one square-degree regions centered on each LORRI dark sky pointing. We use the transformation provided in the {\it Gaia} DR2 photometric validation paper \citep{DR2phot} to convert the {\it Gaia} photometry to the V-band,
\begin{equation}
\begin{split}
    G - V = & -0.01746 - 0.006860(G_{BP} - G_{RP})  \\
    & - 0.1732(G_{BP} - G_{RP})^2.
\end{split}
\label{eq:gphot}
\end{equation}

The agreement between the {\it Gaia} observations and the {\tt TRILEGAL} simulations is excellent. This agreement is, in part, due to the fact that the {\tt TRILEGAL} algorithm has been calibrated against actual star counts, providing confidence that the extrapolation to fainter flux levels is likely to be quite reliable. The average difference between the simulated star counts and the observed {\it Gaia} star counts for our target fields is $5.5$\% over the range $10 \ge V > 19$ mag. The variation in star counts from field-to-field is primarily due to variation in the Galactic coordinates of the fields, as demonstrated in Figure~\ref{fig:svar}.

We generate an estimate of the total V-band surface brightness of undetected stars in the LORRI fields by integrating the simulated star counts from 30th mag to the detection limit of the LORRI images using the expression
\begin{equation}
\begin{split}
    \mu_{*}({\rm mag\ arcsec^{-2}}) = \phantom{999999999999999999999999999999} \\ 
    -2.5\ {\rm log_{10}} \left[ {1\ deg^2 \over (3600^2\ arcsec^2)} \int_{m_{Faint}}^{m_{Lim}} N(m)\ 10^{-0.4m}\  dm\right]
\end{split}
\label{eq:sbint}
\end{equation}
where $N(m)$ is the differential number of stars per unit magnitude per square degree predicted by the {\tt TRILEGAL} model, $dm$ is the magnitude interval used in the integration (0.10 mag), $m_{Faint}$ is 30 mag, and m$_{Lim}$ is the detection limit ($\sim 19.1$ mag in V for the typical LORRI image). We use the $N(m)$ derived from the simulated star catalogs for each of the seven fields. The faint object surface brightness results for each NH dark sky field are presented in Table~\ref{tab:udisb}. The errors for the surface brightnesses of the undetected faint stars are derived from the 1-$\sigma$ $\sqrt{N}$ uncertainties in the simulated {\tt TRILEGAL} number counts. The surface brightness levels in Table~\ref{tab:udisb}
define the normalization for our spectral energy distributions (SEDs) that are used to derive the estimated contribution of undetected sources to the total LORRI signal.  

\begin{deluxetable}{cccccccc}
\tabletypesize{\scriptsize}
\tablecolumns{3}
\tablewidth{0pt}
\tablecaption{Estimated V-band Surface Brightness Levels for Undetected Sources}
\tablehead{\colhead{ }&
\colhead{V-mag}&
\colhead{$\mu_{\rm V}$} \\
\colhead{Data Source}&
\colhead{Integr. range}&
\colhead{mag arcsec$^{-2}$}&}
\startdata
OE394 star sim.   & 19.1 - 30.0 & 29.79 $\pm$ 0.06 \\
OJ394 star sim.   & 19.1 - 30.0 & 29.37 $\pm$ 0.05 \\
n3c61f star sim.  & 19.1 - 30.0 & 29.59 $\pm$ 0.06 \\
ZL138a star sim.  & 19.1 - 30.0 & 29.65 $\pm$ 0.06 \\
ZL138b star sim.  & 19.1 - 30.0 & 29.62 $\pm$ 0.06 \\
ZL138c star sim.  & 19.1 - 30.0 & 29.54 $\pm$ 0.06 \\
ZL138d star sim.  & 19.1 - 30.0 & 29.49 $\pm$ 0.05 \\
\hline
Galaxy Counts     & 19.1 - 30.0 & 27.74 $\pm$ 0.32 \\
\enddata
\tablecomments{The value in column (3) is the integrated V-band surface brightness (equation~\ref{eq:sbint}) over the magnitude range given in column (2). For stars, the number counts are based on the {\tt TRILEGAL} \citep{tril2005, trilv16} Milky Way model simulations for the seven NH dark sky fields. All surface brightness values are on the AB system.}
\label{tab:udisb}
\end{deluxetable}

\begin{figure*}[htbp]
\centering
\includegraphics[keepaspectratio,width=6.5 in]{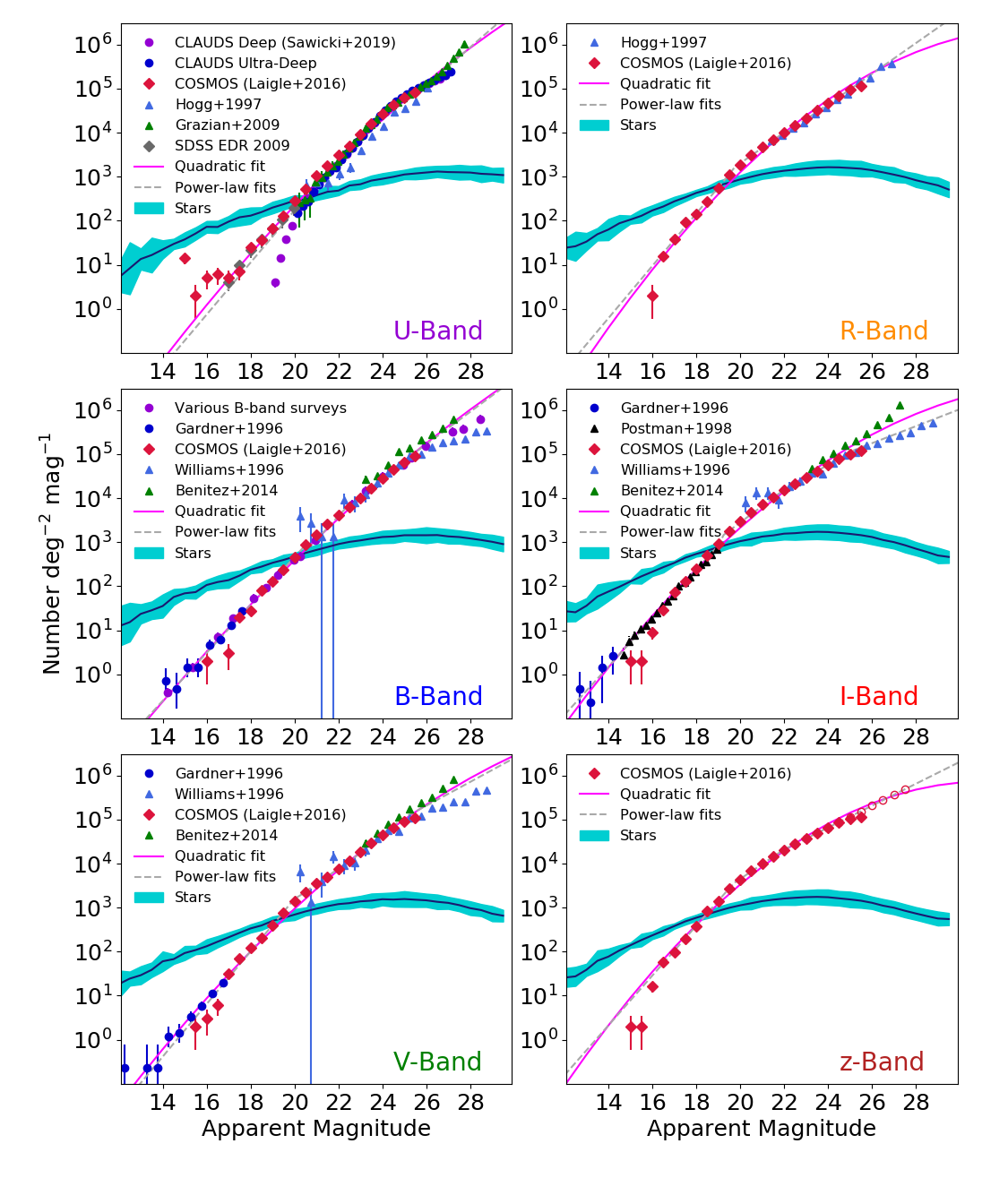}
\caption{The number of galaxies per square degree per magnitude in the UBVRIZ passbands. Also shown are the best fit multi-range power law fits (dashed grey lines) and the best fit quadratic relation (magenta curve). For reference, we also show the star counts for each passband (light blue shaded curve) based on the {\tt TRILEGAL} \citep{tril2005, trilv16} simulations (see \S~\ref{ssec:fstar}). Apparent magnitudes are on the AB system.}
\label{fig:galNmag}
\end{figure*}

The SED of the stellar population varies with limiting magnitude due to the increasing fraction of late-type stars and low-mass stars at fainter magnitudes. Given the very broad passband of LORRI (Figure~\ref{fig:lorribandpass}), it is  essential that we model the change in average stellar SED as a function of magnitude in order to compute a robust estimate of the signal from undetected stars. We estimate the dependence of the SED of the stellar population on apparent magnitude by computing the mean UBVRIJHK photometry for stars in the {\tt TRILEGAL} simulations in eleven 1-magnitude wide bins from V=19 to V=30. In each bin, we use the mean photometry to generate an average SED, $F_\lambda(\lambda)$, for that bin, normalized to the V-band surface brightnesses given in Table~\ref{tab:udisb}. The field-to-field variation in the mean stellar colors as a function of magnitude in the {\tt TRILEGAL} simulations of the NH target regions is negligible ($<0.04$ mag).

For each magnitude bin, we then compute the expected LORRI count rate per unit solid angle, ${\rm C_L}$, for the undetected faint stars by integrating the appropriate SED of the sources over the LORRI response function, where
\begin{equation}
    {\rm C_{L}}({\rm DN\ s^{-1}\ sr^{-1}}) = {1 \over \Omega_{pixel}}\ \int_{\lambda_1}^{\lambda_2} R(\lambda) F_{\lambda}(\lambda) d\lambda.
\label{eq:lorricounts}
\end{equation}
\noindent where $R(\lambda)$ is the LORRI absolute responsivity function (shown in Figure~\ref{fig:lorribandpass}), $F_{\lambda}(\lambda)$ is the SED expressed as a flux density, and $\Omega_{pixel}$ is the solid angle subtended by a single LORRI $4\times4$ binned pixel ($3.91264 \times 10^{-10}$ sr). The flux density, $F_{\lambda}(\lambda)$, for each magnitude bin is generated by fitting a cubic spline to the derived mean UBVRIJHK magnitudes (in the AB system) using a wavelength step size, d$\lambda$, of 10 nm, a LORRI $4\times4$ pixel area of 16.6464 arcsec$^2$, and then converting to the appropriate cgs units (erg cm$^{-2}$ s$^{-1}$ nm$^{-1}$). The integration is performed over the full LORRI sensitivity range from 0.30 $\mu$m to 1.00 $\mu$m. The LORRI counts in a 30 s exposure are then computed and converted to an intensity expressed in units of nW m$^{-2}$ sr$^{-1}$ via  Equation (\ref{eq:s2iconv}). The total signal over the full magnitude range $19 < V \le 30$ is then just the number-count weighted sum of the ${\rm C_L}$ values in each magnitude bin. The intensities for the faint field stars derived in this way are listed in Table~\ref{tab:skyflux}. The turnover in the star counts at magnitudes fainter than V=24 mag (see Figures~\ref{fig:vstars}~and~\ref{fig:galNmag}) means that the derived sky intensity from undetected faint stars does not change significantly so long as the faint limit used in Equation~\ref{eq:sbint} is $V>24$ mag at which point the integral reaches $>98$\% of the value obtained with our chosen integration limit of $V=30$ mag.

The errors in the derived intensities from the {\tt TRILEGAL} simulations are a combination of statistical $\sqrt{N}$ errors in the star counts and systematic variations that depend on model-specific parameters. As demonstrated in both Figures~\ref{fig:vstars}~and~\ref{fig:svar}, the default {\tt TRILEGAL} parameters reproduce the observed {\it Gaia} star counts at ${\rm V} \le 20$ extremely well. But to allow for plausible model uncertainty, we estimate the systematic changes on the derived sky intensity due to changes in the model parameters that shift the derived model star counts by $\pm2\sigma$ from the observed star counts. Shifts of this amplitude are achieved by systematically changing the disk scale heights, halo effective radii, and halo oblateness by $\pm$10\%. Allowing the model parameters to vary over a larger range than this would result in simulations that were in strong disagreement with the observations. Over the range $19 < V \le 30$, the {\tt TRILEGAL} models predict that, for the NH fields, 20\%, 24\%, and 56\% of the stars, respectively, are part of the thin disk, the thick disk, and the halo. Hence, parameters for all three components are key to setting the predicted star counts at high galactic latitudes, albeit at magnitudes fainter than $V=25$, halo stars account for $\sim$68\% of the population. 

In sum, the statistical error in the faint star sky contribution accounts for less than 10\% of the uncertainty in this signal; the error in the sky component due to undetected faint stars is dominated by the systematic uncertainties associated with variation in the {\tt TRILEGAL} models. The combined fractional error from both the statistical and systematic uncertainties is 14\% $-$ 17\% of the derived faint star signal and, furthermore, the signal due to those undetected faint stars contributes, on average, just $\sim$7\% of the total sky signal.

\subsection{Integrated Galaxy Light (IGL)}\label{ssec:fgal}

We estimate the IGL in a manner similar to that used to compute the contribution of faint stars below the LORRI detection limit. We use a compilation of the well-measured galaxy number counts in the UBVRIZ passbands. We take galaxy number counts from various literature sources with incompleteness corrections applied to the raw number counts. Specifically, we use the observed galaxy number counts from the following works:  \cite{hogg97,sdssedr,graz2009,cosmos2016,clauds2019} for the U band, \cite{gardner96,hdf96,ben2004,cosmos2016} for the B and V bands, \cite{hogg97,cosmos2016} for the R band, \cite{gardner96,deeprange98,hdf96,ben2004,cosmos2016} for the I-band, and \cite{cosmos2016} for the z band.

We fit the observed differential counts per unit area and per unit mag, $N(m)$, with both a quadratic curve and with a sequence of four power laws covering the magnitude range 14 to 28. Both functional forms give reasonable representations of the galaxy number counts and are shown along with the observations in Figure~\ref{fig:galNmag} for the UBVRIZ passbands. The IGL, expressed as a V-band surface brightness, is given in Table~\ref{tab:udisb}. Our estimate for the surface brightness of faint galaxies is derived using equation~(\ref{eq:sbint}) with integration limits running from mag=30 to mag=19.1 and we use the best multi-power law fits to the observed, completeness-corrected number counts from various surveys. The error in the galaxy surface brightness limit listed in Table~\ref{tab:udisb} includes the uncertainties in number counts, uncertainties in the best fit parameters, reasonable variations in the form of those best fit functions, and an estimate of the cosmic variance (see next paragraph). The uncertainty in the faint end slope of the log(N)-magnitude relation is the dominant source of error in the derived IGL. For the LORRI images used in this study, the incident light from Milky Way stars below the detection limit is sub-dominant (by a factor of $\sim5$) compared to that from the IGL. 

In contrast with the stellar population, the galaxy population tends towards bluer SEDs at fainter magnitudes. We use the COSMOS multi-band survey \citep{cosmos2016} to generate mean galaxy SEDs in bins 1 magnitude wide from V=19 to V=25 using the UBVRIJHK absolute magnitude data provided in the COSMOS catalog. Below V=25, the catalog is less complete and we just assume the SEDs for galaxies with V$>25$ are identical to those in the $24 < V \le 25$ bin. We then follow the same prescription as discussed in Sec.~\ref{ssec:fstar} using equation~(\ref{eq:lorricounts}) to compute the signal in each bin using the appropriate SED and then generate the final IGL signal from the number-count weighted sum of the ${\rm C_L}$ values for each bin. The IGL intensity derived in this way is listed in Table~\ref{tab:skyflux}. We assume that the IGL signal does not vary significantly between LORRI fields. This is a reasonable assumption since the $\sqrt{N}$ variations in the cumulative galaxy number counts between V=19.1 and V=30 are $< 0.2$\% and the effect of cosmic variance on the galaxy number counts on 0.3$^\circ$ scales down to such depths is estimated to be $\pm 9.4$\% integrated over the redshift range $0 < z \leq 6$ based on the approach presented in \citet{cosvar2008}. The cosmic variance is included in the $\pm 0.32$ mag arcsec$^{-2}$ uncertainty in our surface brightness estimate (Table~\ref{tab:udisb}) and in the total uncertainty of $\pm 2.20~{\rm nW~m^{-2}~sr^{-1}}$ in the derived IGL intensity (Table~\ref{tab:skyflux}). The IGL intensity is only weakly dependent on the integration limit used providing that limit reaches to at least $V = 30$. If we extend the integration of the galaxy number counts to fainter limits, the predicted IGL intensity will increase but the rise depends on the faint-end slope of the galaxy counts (see Figure~\ref{fig:faintendslope}). If the faint-end slope values observed at $V < 29$, which typically lie in the range $0.25 - 0.35$, continue to $V > 30$ then the derived IGL intensity would increase only by 4\% $-$ 13\% even if we extend the integration down to $V = 34$. Such an increase is covered by the $\sim30$\% errors on our present IGL estimate. Further discussion of this issue is found in $\S\ref{sec:igldiscuss}$.

\subsection{Scattered Starlight and Galaxy-Light}

\begin{figure}[htbp]
\centering
\includegraphics[keepaspectratio,width=3.5 in]{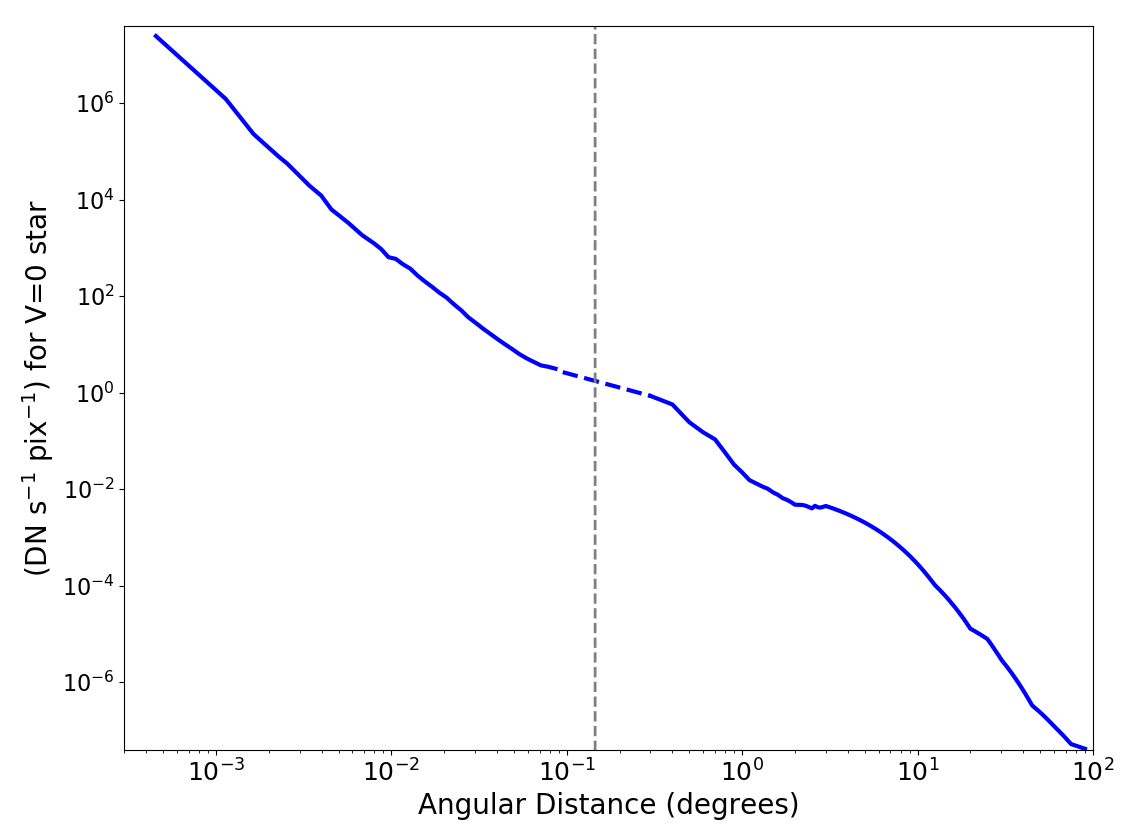}
\caption{The large-scale PSF for LORRI expressed in counts per pixel accumulated in 1 second for a V=0 magnitude star as a function angular distance. The rates are for the $4\times4$ pixel binning mode. The dashed vertical line denotes the half-width of the LORRI detector footprint on the sky. Data to the left of this line is obtained from in-flight images of stars, data to the right is obtained from pre-flight testing and from in-flight observations of scattered sunlight.  The dashed-line part of the PSF profile is an interpolation over a radial-zone with no data.}
\label{fig:LORRIPSF}
\end{figure}

LORRI is sensitive to sunlight scattered into the detector even out to solar elongations of $90^\circ,$ beyond which the LORRI aperture is in  the spacecraft's shadow \citep{lorri, lorri2}. This implies that scattered {\it starlight} may also be an important contribution to the total sky level, as is demonstrated in the next section. Figure \ref{fig:LORRIPSF} shows the complete composite PSF/Scattering-function for LORRI, which describes the radial distribution of scattered light from the pixel-scale to large angles.  The inner part  of the function was determined by LORRI images of stars in an open cluster, augmented with exposures of Vega and Arcturus to extend the PSF out to the angular limits of the LORRI field of view.  Pre-launch calibration extends the distribution to the few-degree scale (with small gap that is interpolated over). The large-angle portion of the scattering function was based on the amplitude of scattered sunlight as a function of solar elongation, where the light intensity at any angle was taken as the median flux level over the full LORRI field.  We allocate a 10\%\ uncertainty for the PSF amplitude at any location, based on observed variations in the scattered-light amplitude with azimuth at a constant solar-elongation and fine-scale noise in the profile as a function of radius. In passing we note that the LORRI photometric calibration, which we use throughout this paper, is established by integrating the light of standard stars within a 10 arcsecond aperture on the CCD.  Integrating the PSF over the full sky yields a total flux 7.8\% larger that the integral flux within this aperture.

\subsubsection{Estimation of the Scattered Starlight Contribution}\label{sec:ssl}

\begin{figure*}[htbp]
\centering
\includegraphics[keepaspectratio,width=6 in]{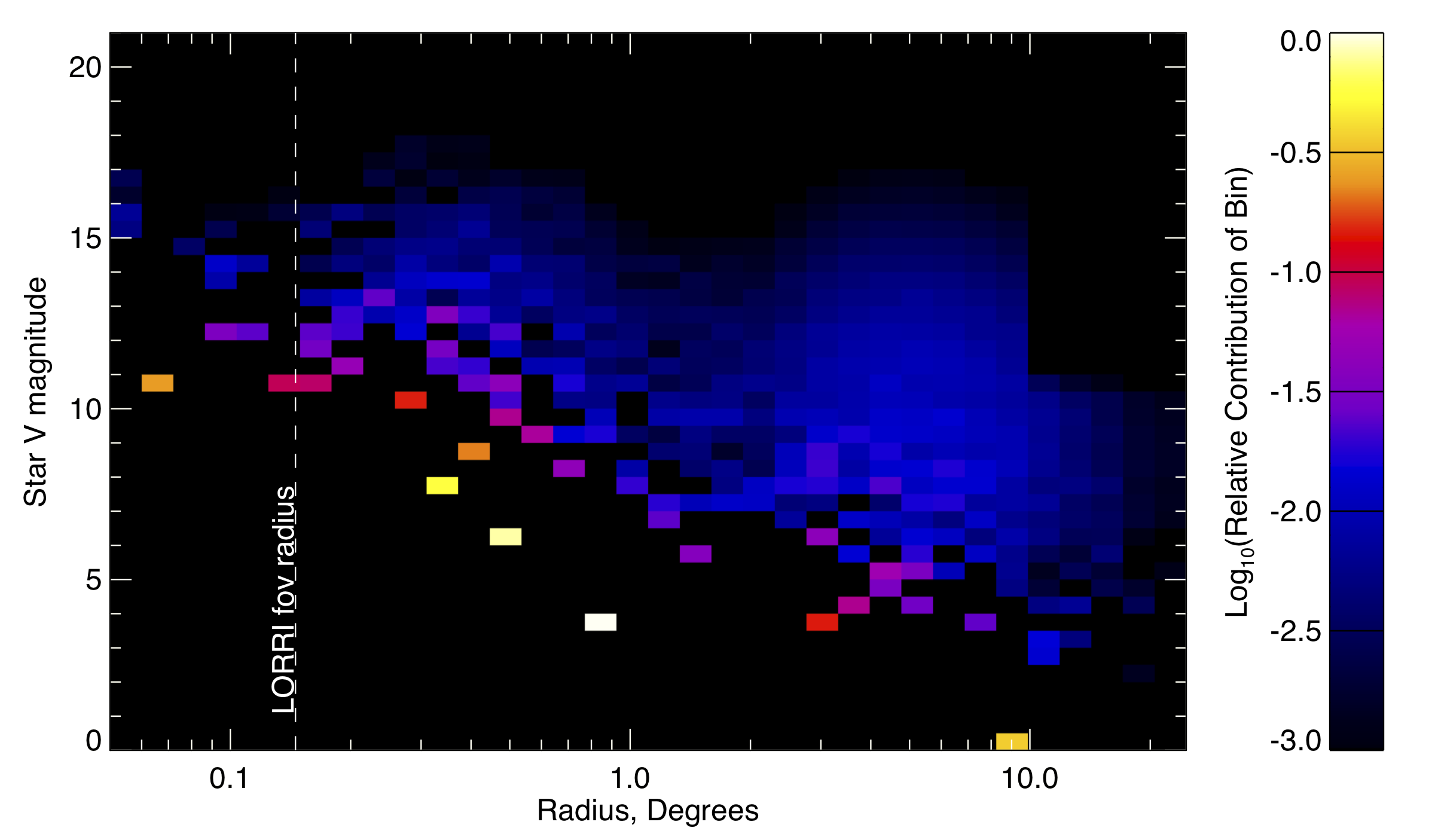}
\caption{The relative contribution of scattered starlight to the detected sky level as a function of the stellar magnitude and the angular separation between the star and the field center of the LORRI camera. The data in this figure are for the n3c61f field and represent data from a combined sample generated from the {\it Gaia} DR2, Tycho2, and Yale Bright Star v5.0 catalogs. Beyond $10^\circ$ only the stars brighter than $V \sim 11$ contribute significantly to the scattered light signal.}
\label{fig:SSLcontours}
\end{figure*}

The first step in estimating the scattered starlight (SSL) contribution was to assess the maximum angular scale we need to include in our calculations. While Figure \ref{fig:LORRIPSF} indicates scattered light can be seen as much as $90^\circ$ off-axis, the amplitude of the signal is decreasing rapidly with increasing angular distance from the field center. An initial assessment of each NH field indicated that we would need to account for scattered light from stars as far as $20^\circ$ to $45^\circ$ from the field center. This is highlighted in Figure~\ref{fig:SSLcontours}, which shows the relative contribution of scattered light for stars in the vicinity of the n3c61f field.  We find that in our seven fields the stars with angular separations in the range $10^\circ$ to $20^\circ$ contribute up to 6\% of the total scattered starlight signal. Stars in the range $20^{\circ}$ to $45^\circ$ contribute no more than 1.3\% of the total scattered starlight signal. Beyond $45^\circ$ even the brightest known stars contribute negligibly ($<0.001$\%) to the detected sky level.

The next step was to assess how faint and how complete a star catalog we would require. The most robust star catalog currently available is the {\it Gaia} DR2 catalog \citep{gaia2016,gaia2018} which is complete over the entire sky down to $V \sim 20$ mag. Extracting stars to this depth out to $45^\circ$ for each field would involve a very large number of sources ($\sim13$ million stars for each field in our study). Hence, we performed a test to see if we could use stars to this depth out to a smaller angular scale and supplement that list with a shallower catalog out to the full $45^\circ$. We used the ZL138b field as a test case. We extracted all {\it Gaia} stars down to $V = 19.5$ out to an angular separation of $20^\circ$, supplemented with any missing bright stars ($2 \leq V \leq 8$) from the Tycho2 star catalog \citep{tycho2} and, for ${\rm V} < 2,$ from the Yale Bright Star catalog v5.0 \citep{YBSC5}. We computed the scattered light signal (details given below) and compared that with the scattered light signal derived from a sample with {\it Gaia} stars down to the same depth but only extending out to $10^\circ$ and supplemented with the Tycho2 catalog down to $V = 10.75$ over the angular separation range $10^\circ < \theta \leq 45^\circ$. The scattered light signal from the 20-deg deep {\it Gaia}-Tycho2-YBSC5 sample and the 45-deg {\it Gaia}-Tycho2-YBSC5 catalog agree to within 0.35\%. We thus conclude the hybrid catalogs (deep {\it Gaia} data out to $10^\circ$ supplemented with shallower Tycho2 and Yale bright star data) would allow for a robust estimate of the scattered light signal and significantly reduce the amount of star data needed (about 730,000 stars per field instead of $>10$ million). Our adopted procedure is described in detail below.

To perform the scattered starlight estimation, we identify the stars for each field with $V \le 19$ that lie within an angular distance of $\theta \le 10^\circ$ from the mean LORRI pointing (see Table~\ref{tab:obs}). We supplement this list with all stars with $V \le 10.75$ that lie between $10^\circ < \theta \le 45^\circ$ from each field center.  We use the {\it Gaia} DR2 catalog to identify stars with $\theta \le 10^\circ$ and $8 < V \le 19$. To ensure we have complete coverage of all known bright stars, we use the Tycho2 catalog to identify stars with $\theta \le 10^\circ$ and $2 \le V \le 8$ or with $10^\circ < \theta \le 45^\circ$ and $2 \le V \le 10.75$. We merge the data from the two catalogs with care to reject common objects. We use the established flux transformations to convert the {\it Gaia} and Tycho2 photometry to the Johnson V system. Any Tycho2 and {\it Gaia} DR2 entries that lie within $3''$ of one another and which have an apparent magnitude difference of $\Delta V \le 0.20$ mag are considered to be duplicate entries. For any duplicate entries, we adopt the {\it Gaia} DR2 values for the position and flux. Lastly, we use the Yale Bright Star catalog v5.0 to include stars with $V < 2$ and $\theta \le 45^\circ$. This procedure generates a star catalog for each field that is complete to $V = 19$ out to an angular distance of $10^\circ$ from the field center and to $V = 10.75$ for angular distances from $10^\circ$ to $45^\circ$. We also derive an estimate of the (B-V) color for each star using the published color transformations. For {\it Gaia} data, we use the DR2 transformation to compute (V-R) and then use the {\tt TRILEGAL} simulations to derive the transformation from (V-R) to (B-V). The total uncertainty in our (B-V) colors is $\sim0.1$ mag.

For stars within 1.5 LORRI field-widths of the center ($26\amin1$) we account for the SSL variation over the field as well as for the faint extended wings of stars {\it within} the fields that extended beyond the $12''$ masking radius applied to the cores of the stellar PSFs. For this inner circle around each field center, we constructed an image of the star field from the combined star catalogs and convolved it with the composite PSF (Figure~\ref{fig:LORRIPSF}), and then measured the integrated SSL contribution over the field. For our adopted photometric detection limit of 19.1 V mag, all the stars masked within the fields are included in the {\it Gaia} catalog. The SSL value due to stars within 1.5 LORRI field-widths is determined from the mode of the pixel intensity histogram in the above convolved image, with star centers masked out using the same process as was done for the observations. This procedure accurately allows us to account for any gradients in the SSL across the LORRI field of view. Color corrections, needed due to the broad wavelength response of the LORRI instrument, were applied for stars with different (B-V) values using the methodology discussed below.

For stars with angular separations larger than 1.5 LORRI field-widths, we simply calculate the scattered light contribution of any star relative to its separation from the center of the LORRI field under the assumption that any SSL gradients across the LORRI field of view from these more angularly distant stars will be negligible.\footnote{For the three DKBO fields, the SSL contribution is calculated for each position in the sequence.}
To compute the SSL contribution for stars with angular separation larger than 1.5 LORRI field-widths, 
we estimate the total V-band surface brightness reaching the central LORRI $4\times4$ pixel due to scattered starlight as a function of the (B-V) color, $\mu_{V,SSL}(B-V)$, using the following expression
\begin{equation}
\begin{split}
    \mu_{V,SSL} (B-V) = \phantom{99999999999999999999999999999999} \\
    -2.5 {\rm log_{10}} \left[ {1 \over (4.08'')^2} \sum_{i=1}^{i=N} {\rm PSF(\theta)\  10^{-0.4V}} \right] + 18.88
    \label{eq:psfcounts}
\end{split}
\end{equation}
\noindent where the sum is over all stars within a given (B-V) range, PSF$(\theta)$ is the value of the composite PSF for an angular distance, $\theta$, between the star and the LORRI field center, V is the V-band magnitude of the star, and where 18.88 is the appropriate zeropoint for a $4\times4$ binned LORRI pixel \citep{lorri2}. We compute the value of $\mu_{V,SSL}(B-V)$ for five (B-V) color ranges: [1] $(B-V) < 0.00$, [2] $0.00 \le (B-V) < 0.25$, [3] $0.25 \le (B-V) < 0.50$, [4] $0.50 \le (B-V) < 1.00$, and [5] $(B-V) \ge 1.00$. We use the {\tt TRILEGAL} simulations to generate SEDs for these same five color ranges for stars with $5 \le V \le 19$. We renormalize these SEDs to match the V-band surface brightnesses computed using equation (\ref{eq:psfcounts}) and then use equation  (\ref{eq:lorricounts}) with the appropriate renormalized SED to compute the predicted LORRI count rates for each (B-V) range. The final step is to sum up the results for all five color bins to get the total predicted scattered starlight signal for each field. This procedure allows us to account for the sensitivity of the broad LORRI response function to stars with significantly different SEDs. Our estimated scattered starlight signal for each field is listed in Table~\ref{tab:skyflux}.  The uncertainty in the estimated scattered starlight signal is 10\%, due to the uncertainty in the determination of the composite PSF.

\subsubsection{Estimation of the Scattered Galaxy-Light Contribution}

Scattered light from bright ($V \le 19$) galaxies outside of the LORRI field of view can also be computed but is expected to be negligible. To make the scattered galaxy light estimate, we first compute the mean V-band surface brightness for bright galaxies in the range $10 \le V \le 19$, $\mu_{V,BG}$, using equation (\ref{eq:sbint}). We then compute the surface brightness of scattered light from this galaxy population using the expression:
\begin{equation}
\begin{split}
    \mu_{V,SGL} =  \phantom{999999999999999999999999999999999999999} \\  
    -2.5 {\rm log_{10}} \left[  \sum_{\theta=0.03^\circ}^{\theta=45^\circ} {\rm PSF(\theta)\  [\pi (\theta_{i+1}^2 - \theta_{i-1}^2)]\ 10^{-0.4\mu_{V,BG}}} \right] \\
    + 18.88
    \label{eq:scatgal}
\end{split}
\end{equation}
\noindent where the sum is over angular separations from 0.03$^\circ$ to 45$^\circ$ in annuli $0.02^\circ$ in width and the other expressions are the same as defined in equation  (\ref{eq:psfcounts}). We derive the SED for galaxies in this magnitude range using the UBVRIZYJHK data from the COSMOS survey \citep{cosmos2016} and renormalize this SED to match the $\mu_{V,SGL}$ surface brightness derived with equation (\ref{eq:scatgal}). We then use equation (\ref{eq:lorricounts}) to derive the predicted signal from scattered light from galaxies outside the LORRI fields. The results are presented in Table~\ref{tab:skyflux}. As anticipated, the contribution to the total scattered light from galaxies is at least $\sim100$ times smaller than that from stars.

\subsection{Diffuse Galactic Light (DGL)}\label{sec:dgl}

The final optical background correction that we apply is to account for Milky Way starlight scattered by interstellar dust into our line of sight.  Following \citet{zemcov}, this ``diffuse galactic light," or DGL component, is estimated for any given LORRI field  from, $I_{100},$ the strength of the $100~{\rm\mu m}$ thermal emission of the infrared ``cirrus" present in the field. The ``IRIS" reprocessing of the IRAS full-sky thermal-IR maps \citep{iris} provides the needed $100~{\rm\mu m}$ measures. The conversion between $I_{100}$ and DGL assumes a linear scaling between the two, which should be valid for modest dust optical depth. 

This simple proscription, however, belies a number of uncertainties in how it is actually applied to provide an accurate quantitative estimate of DGL, given $I_{100}.$  We examined variety of estimators, as well as searching for systematic biases in the input $I_{100}$ maps.  Our initial approach was to evaluate the  approach of \citet{zemcov}, who attempted to derive a DGL estimator based on the theoretical scattering properties of plausible models of the dust grains.
Allowing for a slowly varying phase function dependent on galactic latitude, they give the DGL scattered-light signal in the LORRI-passband as
\begin{equation}
I_{DGL}=C_{100}~{I_{100}\over 1{\rm~MJy~sr^{-1}}}~{1-0.67\sqrt{\rm sin|b|}\over0.376}~{\rm ~nW~m^{-2}~sr^{-1}}
\label{eqn:conv}
\end{equation}
where the term on the right gives the dependence of the conversion on galactic latitude, b, and is normalized to  be  unity  at ${\rm b=60^\circ}.$  \citet{zemcov} note that it is important to subtract $0.78{\rm~MJy~sr^{-1}}$ from the IRIS flux values to correct for the contribution of the cosmic infrared background (CIB) \citep{cib1, cib2}. The specific value of the leading conversion constant that they derive  is $C_{100}=9.8\pm3.9.$\footnote{Technically, \citet{zemcov} provides a coefficient that converts $\nu I_\nu$ calculated from the $100~{\rm\mu m}$ background maps to the needed $\lambda I_\lambda$ units of the DGL.  We cast their coefficient as a $C_{100}$ value specific to $\nu=100~{\rm\mu m},$ that converts the maps' $I_\nu$  directly.}

The nearly $\sim40\%$ relative error in the coefficient, however, translates to a large source of uncertainty in our final COB measurements, thus we considered other approaches to estimating DGL. Intriguingly, \citet{BD2012} provide a direct {\it observational} measurement of the total DGL flux as a function of wavelength and galactic latitude, based on correlating residuals in the Sloan  Digital Sky Survey background spectra with $100~{\rm\mu m}$ flux. Conveniently, they present their DGL spectra as a scaling between optical and $100~{\rm\mu m}$ flux over a wide optical band pass that encompasses the LORRI band pass.  Calculating a LORRI-specific conversion-scaling requires only integrating the LORRI response function over their DGL spectra. Using the \citet{BD2012} $50^\circ<|b|<90^\circ$ DGL spectrum yields $C_{100}=5.5\pm1.3,$  after rescaling the spectrum by their recommended $2.1\pm0.4$ bias correction factor.  DGL estimates produced by this methodology have both markedly smaller amplitudes and errors than does those provided by the \citet{zemcov} estimator.  That said, unfortunately, the particular high-galactic latitude spectrum may require an even larger bias correction, given the small number of observations used to generate it (Brandt, private communication).  It is thus likely that this methodology underestimates the appropriate DGL correction. Its nominally smaller formal errors do not  reflect this potential systematic error.

In passing, we note that we also attempted to estimate DGL corrections directly from our data by correlating the residual sky level after all correction besides the DGL correction have been applied with $100~{\rm\mu m}$ flux. Unfortunately, given the amplitude of the errors in the residual sky levels and the small range of $100~{\rm\mu m}$ flux over our sample, we could not derive a significant scaling coefficient. As emphasized by \citet{zemcov}, this approach may be useful for future work with a larger sample of fields, but we will not pursue this further in the present work.

\begin{figure*}[htbp]
\centering
\includegraphics[keepaspectratio,width=6.2 in]{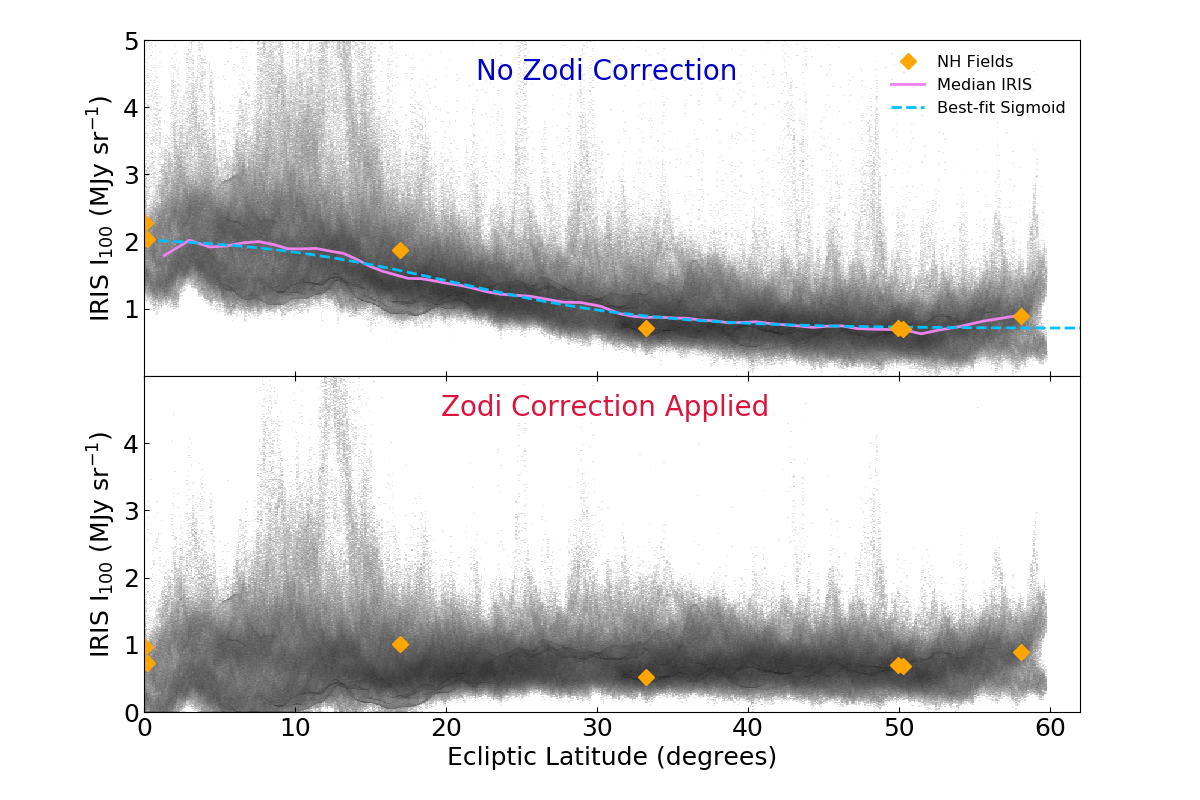}
\caption{The upper panel shows the IRIS $100~{\rm\mu m}$ flux values for high galactic latitude regions of sky with $|b|\geq60^\circ$ and $55^\circ \leq l \leq 355^\circ$ as a function of absolute ecliptic latitude. The images show the logarithmic density distribution of pixels in the IRIS map. The IRIS $100~{\rm\mu m}$ flux data were kindly provided to us by C. Bot (private comm.). Orange symbols indicate the $100~{\rm\mu m}$ fluxes of the seven New Horizons fields. A trend of increasing flux with decreasing ecliptic latitude is clearly evident, indicative of an incomplete zodiacal light correction. The median $100~{\rm\mu m}$ flux at any ecliptic latitude (violet line) is fitted with an inverse sigmoid function (dashed blue line).  The bottom panel shows the $100~{\rm\mu m}$ and New Horizon fluxes after a correction derived from this fit is subtracted.}
\label{fig:iris}
\end{figure*}

\subsubsection{Measurement of the $100~{\rm\mu m}$ fluxes for the LORRI Fields}

The IRIS $100~{\rm\mu m}$ fluxes for the seven LORRI fields are given in Table \ref{tab:obs}. The fluxes were calculated as an average over the LORRI field of view for the zodiacal-light sequences, and a slightly larger area for the DKBO sequences to accommodate the shifting positions of the images needed to track the DKBOs. The dispersions in the map values for all fields were negligible compared to other sources of error.

In attempting to evaluate the quality of our $100~{\rm\mu m}$ fluxes we compared the IRIS measures to those of \citet{sf11}.  The two sources agreed well for the fields  at high ecliptic-latitudes, but not for two of the fields at low ecliptic latitudes. The IRIS maps indicated markedly stronger $100~{\rm\mu m}$ fluxes, and the implied IRIS-based DGL-corrections yielded residual sky fluxes with a larger variance over the sample than that measured prior to the correction.   Inspection of the IRIS maps showed artifacts associated with imperfect correction for zodiacal light close to the ecliptic plane, however.  By plotting all IRIS $100~{\rm\mu m}$ flux values in the maps at galactic latitude $|b|\geq60^\circ$ as a function of absolute ecliptic latitude (Figure \ref{fig:iris}), a trend indicative of residual zodiacal dust at low ecliptic latitudes is evident. The \citet{sf11} maps show a qualitatively similar trend, but of lower amplitude.

We compute a correction for the residual zodiacal light component in the IRIS data by first finding the median fluxes in $1^{\circ}$--wide ecliptic latitude bins running from $0^\circ \le |\beta| \le 60^\circ$ for the 5.6 million pixels in the IRIS $100~{\rm\mu m}$ sky map with galactic latitude $|b|\geq60^\circ$ and galactic longitude $55^\circ \le l \le 355^\circ$ shown in Figure \ref{fig:iris}. The IRIS $100~{\rm\mu m}$ flux and median data are provided to us by C. Bot (priv. comm.). A constant CIB level of 0.78 MJy sr$^{-1}$ \citep{cib1, cib2} has been subtracted from all IRIS flux values. We then fit an inverse sigmoid function to the median flux vs ecliptic latitude data of the form:
\begin{equation}
    f_{100}(|\beta|) = f_{max} - {(f_{max}-f_{min}) \over (1. + ({\rm e}^{\kappa (|\beta| - \beta_{0})})^{\gamma})}
    \label{eq:sigmoid}
\end{equation}
where $f_{100}(|\beta|)$ is the predicted median $100~{\rm\mu m}$ flux in MJy sr$^{-1}$ at ecliptic latitude, $\beta$, in degrees. The best fit sigmoid parameters are $f_{min} = 0.71$ MJy sr$^{-1}$, $f_{max} = 2.09$ MJy sr$^{-1}$, $\kappa = 0.20$, $\beta_{0} = 20.40^\circ$, and $\gamma = -0.73$. The best fit is shown in Figure \ref{fig:iris} as the blue dashed curve in the upper plot. The correction for the excess zodi signal in the IRIS data is then just $f_{100}(|\beta|) - f_{min}$, which is subtracted from our measured fluxes. The motivation for using a sigmoid function to fit this trend was simply because it nicely models a continuous transition between two constant levels. There may well be alternative functions that could be used to model the observed trend. The corrected IRIS fluxes are given in  Table \ref{tab:obs}, and are plotted in Figure \ref{fig:iris}.

\begin{figure*}[htbp]
\centering
\includegraphics[keepaspectratio,width=6.2 in]{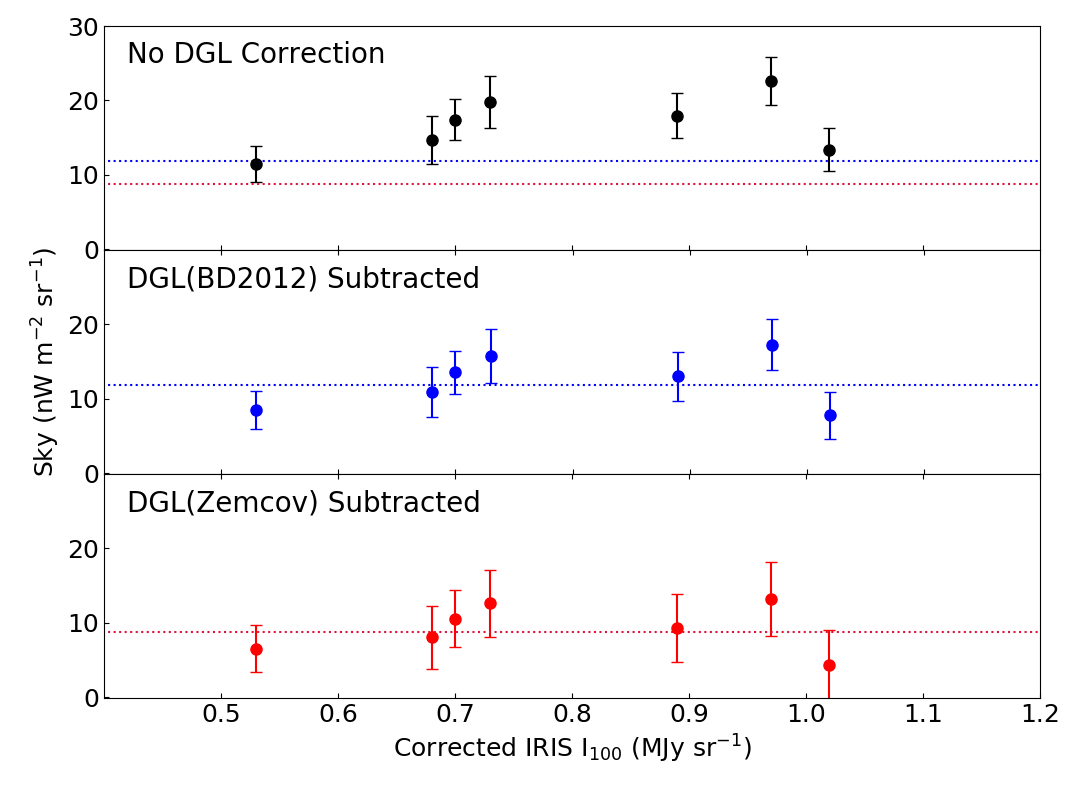}
\caption{The sky levels are plotted before and after subtraction of two different estimates of diffuse galactic light (DGL) components as a function of the average $100~\mu{\rm m}$ flux within each field, based on the IRIS \citep{iris} maps as corrected for residual zodiacal light. All sky levels have also been corrected for light from stars and external galaxies.  The upper panel shows the sky levels prior to the DGL correction. The middle panel shows the final residual levels after the \citet{BD2012} DGL correction is applied. The bottom panel shows the final residual levels after the \citet{zemcov} DGL correction is applied. The points in the bottom two panels show the estimated dCOB level for each field. The final residual sky level, indicated by the horizontal dotted-lines in all panels, is the weighted average dCOB level of all fields for the given DGL estimator.}
\label{fig:sky_val}
\end{figure*}

\begin{deluxetable*}{r|r|r|r|r|r|r|r}
\tabletypesize{\scriptsize}
\tablecolumns{8}
\tablewidth{0pt}
\tablecaption{Sky Flux Decomposition}
\tablehead{
\multicolumn{1}{c|}{}&\multicolumn{7}{c}{}\\
\multicolumn{1}{c|}{}
&\multicolumn{7}{c}{NH Field ID}\\
\multicolumn{1}{c|}{Parameter}
&\multicolumn{1}{c}{OE394 }
&\multicolumn{1}{c}{OJ394 }
&\multicolumn{1}{c}{n3c61f}
&\multicolumn{1}{c}{ZL138a}
&\multicolumn{1}{c}{ZL138b}
&\multicolumn{1}{c}{ZL138c}
&\multicolumn{1}{c}{ZL138d}}
\startdata
{ } & \multicolumn{7}{c}{Total Sky}\\
           Value &  37.52 &  30.34 &  39.75 &  36.48 &  39.80 &  31.73 &  35.69  \\
    Random Error &   1.24 &   0.59 &   1.68 &   2.62 &   2.18 &   2.23 &   1.39  \\
\hline
{ } & \multicolumn{7}{c}{Scattered Light from Bright Field Stars}\\
           Value &  14.74 &   8.58 &  12.03 &   7.03 &   7.43 &   7.16 &   8.29  \\
Systematic Error &   1.47 &   0.86 &   1.20 &   0.70 &   0.74 &   0.72 &   0.83  \\
\hline
{ } & \multicolumn{7}{c}{Integrated Faint Starlight}\\
           Value &   1.95 &   2.87 &   2.34 &   2.22 &   2.28 &   2.46 &   2.56  \\
    Random Error &   0.11 &   0.14 &   0.12 &   0.12 &   0.12 &   0.12 &   0.13  \\
Systematic Error &   0.32 &   0.38 &   0.34 &   0.34 &   0.34 &   0.35 &   0.36  \\
\hline
{ } & \multicolumn{7}{c}{Integrated Faint Galaxy Light}\\
           Value &   7.37 &   7.37 &   7.37 &   7.37 &   7.37 &   7.37 &   7.37  \\
    Random Error &   0.81 &   0.81 &   0.81 &   0.81 &   0.81 &   0.81 &   0.81  \\
Systematic Error &   2.05 &   2.05 &   2.05 &   2.05 &   2.05 &   2.05 &   2.05  \\
\hline
{ } & \multicolumn{7}{c}{Scattered Light from Bright Field Galaxies}\\
           Value &   0.07 &   0.07 &   0.07 &   0.07 &   0.07 &   0.07 &   0.07  \\
Systematic Error &   0.01 &   0.01 &   0.01 &   0.01 &   0.01 &   0.01 &   0.01  \\
\hline
{ } & \multicolumn{7}{c}{Total Sky $-$ (Star + Galaxy Components)}\\
           Value &  13.39 &  11.45 &  17.94 &  19.79 &  22.65 &  14.67 &  17.40  \\
    Random Error &   1.49 &   1.01 &   1.87 &   2.74 &   2.33 &   2.38 &   1.61  \\
Systematic Error &   2.54 &   2.25 &   2.40 &   2.19 &   2.20 &   2.20 &   2.24  \\
     Total Error &   2.94 &   2.47 &   3.04 &   3.51 &   3.20 &   3.24 &   2.76  \\
\hline
{ } & \multicolumn{7}{c}{DGL (from Zemcov)}\\
           Value &   9.00 &   4.96 &   8.62 &   7.19 &   9.47 &   6.63 &   6.82  \\
Systematic Error &   3.58 &   1.97 &   3.43 &   2.86 &   3.77 &   2.64 &   2.72  \\
\hline
{ } & \multicolumn{7}{c}{Residual Sky (w/Zemcov DGL)}\\
           Value &   4.39 &   6.49 &   9.32 &  12.60 &  13.18 &   8.04 &  10.58  \\
    Random Error &   1.49 &   1.01 &   1.87 &   2.74 &   2.33 &   2.38 &   1.61  \\
Systematic Error &   4.39 &   2.99 &   4.18 &   3.60 &   4.37 &   3.43 &   3.52  \\
     Total Error &   4.63 &   3.16 &   4.58 &   4.53 &   4.95 &   4.18 &   3.87  \\
\hline
{ } & \multicolumn{7}{c}{DGL (from BD2012)}\\
           Value &   5.62 &   2.91 &   4.93 &   4.05 &   5.35 &   3.78 &   3.86  \\
Systematic Error &   1.27 &   0.66 &   1.11 &   0.92 &   1.21 &   0.85 &   0.87  \\
\hline
{ } & \multicolumn{7}{c}{Residual Sky (w/BD2012 DGL)}\\
           Value &   7.77 &   8.54 &  13.01 &  15.74 &  17.30 &  10.89 &  13.54  \\
    Random Error &   1.49 &   1.01 &   1.87 &   2.74 &   2.33 &   2.38 &   1.61  \\
Systematic Error &   2.84 &   2.35 &   2.64 &   2.37 &   2.51 &   2.36 &   2.40  \\
     Total Error &   3.20 &   2.56 &   3.24 &   3.63 &   3.43 &   3.35 &   2.89  \\
\hline
\enddata
\tablecomments{All fluxes are in units of ${\rm~nW~m^{-2}~sr^{-1}}.$ The 1$\sigma$ random and systematic
uncertainties in the fluxes are shown below each value, providing they are $\ge 0.005$. For key combined quantities, such as sky residuals, we also give the total error as the quadrature sum of the random and systematic errors. Top row gives the partial program name, followed by ten sections: (1) the total sky intensity, $\lambda I_\lambda$, from Table \ref{tab:skylevs}, (2) scattered starlight flux from stars with $V<19.1,$ (3) estimated flux from stars with $V\ge19.1$ in the LORRI field, (4) estimated flux from galaxies with $V\ge19.1$ in the LORRI field, (5) scattered light from galaxies with $V<19.1,$ (6) Sky flux with star and galaxy fluxes subtracted, (7) Zemcov-based diffuse galaxy light (DGL) background estimated from the 100 $\mu$m flux, (8) Final residual sky signal with all known sources subtracted using Zemcov DGL. (9) DGL based on \cite{BD2012}, (10) Final residual sky with all known sources subtracted using the BD2012 DGL estimate.}
\label{tab:skyflux}
\end{deluxetable*}

We calculated the DGL corrections based on both the \citet{zemcov} and \citet{BD2012} conversion coefficients (Table \ref{tab:skyflux}), given the corrected IRIS $100~{\rm\mu m}$ fluxes. Figure \ref{fig:sky_val} shows the sky levels as a function of  the average $100~{\rm\mu m}$ flux after the total star and galaxy light contributions have been subtracted, before and after the two sets of DGL components have been subtracted.  The final residual sky levels after the DGL subtraction represent estimates of the dCOB in each field. The range in the final dCOB values serves to emphasize the likely effects of uncertainty in the  DGL correction.

\subsection{Sunlight Scattered by Dust Within the Kuiper Belt}\label{sec:ipd}

We did not expect to detect any sunlight scattered by interplanetary dust (IPD) within the Kuiper Belt, based on detailed models of the distribution of dust within the outer solar system \citep{zem18, poppe}, normalized by {\it in situ} measurements by the New Horizons Student Dust Counter \citep{sdc}. In brief, the models integrate the scattered light along any line of sight from the New Horizons spacecraft, using an appropriate phase-function at any given solar elongation for an assumed dust-particle size/density function. For example, the predicted scattered sunlight flux at 40 AU and $90^\circ$ solar elongation is $\sim 0.1{\rm ~nW~m^{-2}~sr^{-1}},$ well over an order of magnitude below our present sensitivity limit (Figure \ref{fig:dust}). We do not include this small term in our analysis.

Despite these arguments, as noted in $\S\ref{sec:obs}$ the  motivation for observing the ``ZodiacLight" fields was a simple test to see if there was any evidence for light scattered by dust within the Kuiper Belt independent of any models. The weighted-average residual-skies of the two zero ecliptic-latitude fields (OE394 was excluded, given its intermediate ecliptic latitude) less that of the three highest ecliptic-latitude fields\footnote{At the  time of  the observations, New Horizons was 1.5 AU above the ecliptic plane, so the zero-latitude fields correspond to sight-lines parallel to the plane with this offset.} is $3.5\pm2.2{\rm ~nW~m^{-2}~sr^{-1}}$ for the Zemcov DGL and $4.0\pm2.1{\rm ~nW~m^{-2}~sr^{-1}},$ assuming the BD2012 DGL.

An IPD measure at this low level of significance is of potential interest; however, it is highly sensitive to our assumed ZL correction to the $100~{\rm\mu m}$ flux maps.  Prior to the ZL correction, we obtained an IPD scattered-light level consistent with zero; in short, reducing the DGL attributed to the low ecliptic-latitude fields implies more residual signal to be accounted for.  Paradoxically, correcting the low ecliptic-latitude fields for this IPD signal {\it increases} the significance of the dCOB level, which will be discussed in the next section by bringing the these fields into better concordance with the remaining sample, thus reducing the amount of random variance encompassed by the full sample.  We thus do not include this potential IPD correction in our dCOB analysis.

\section{A Tentative Detection of a Diffuse Cosmic Optical Background}
\label{sec:detect}

\begin{figure*}[htbp]
\centering
\includegraphics[keepaspectratio,width=6.0 in]{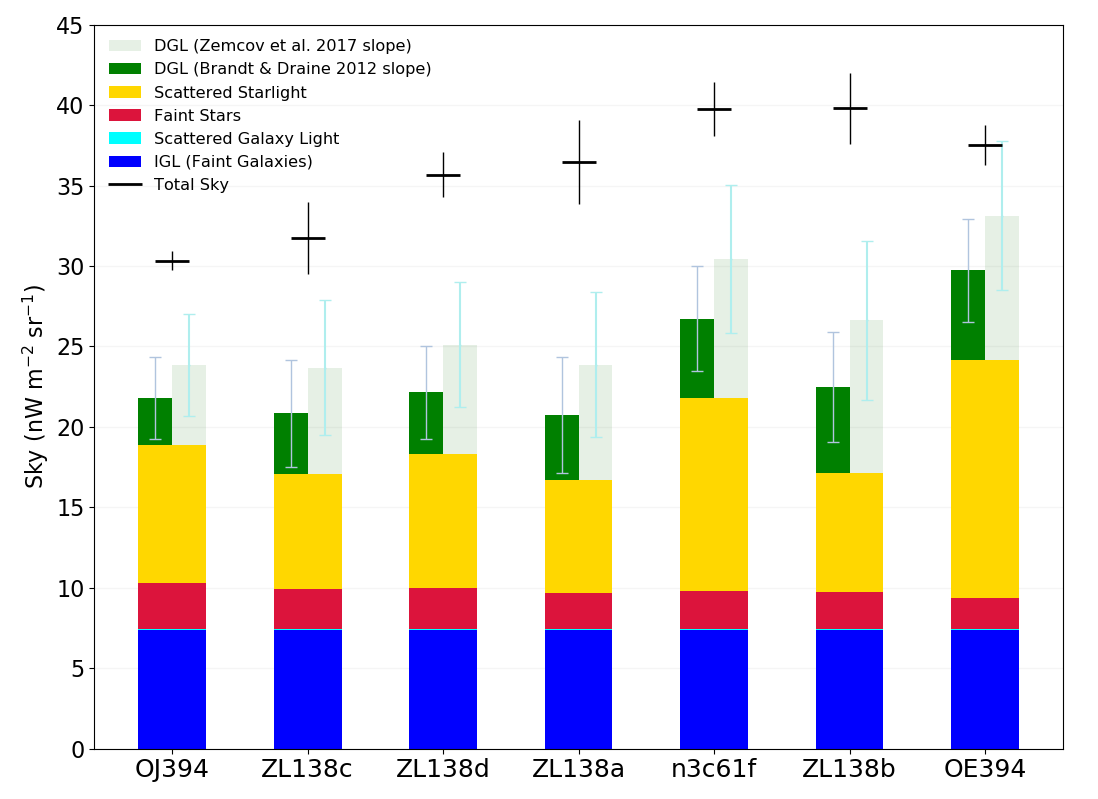}
\caption{A stacked bar chart showing the amplitudes of the known sky components for each of the seven fields observed with LORRI. The black horizontal lines with error-bars show our measured total sky values and their uncertainties for each field.  The fields are sorted in order of increasing $100~{\rm \mu m}$ flux density. The gap between the total-sky levels and the tops of the dark green bars indicates the dCOB component for each field using the DGL based on \cite{BD2012}.
The grey error bars show the quadrature sum of the uncertainties in the known sky components using this estimate for the DGL (i.e., faint stars + IGL + scattered light + DGL). The gap between the total-sky levels and the tops of the light green bars show the results if we compute the DGL using the $100~{\rm \mu m}$ to optical flux relation in \cite{zemcov}. The light blue error bars show the quadrature sum of the uncertainties in the known sky components using this latter DGL estimate.
}
\label{fig:skybars}
\end{figure*}

Figure \ref{fig:skybars} summarizes graphically the estimated components contributing to the total sky measurement for each field.  All seven fields have unaccounted for excess signal above the 2-$\sigma$ level with the \citet{BD2012} DGL correction, as do most of the field with the \citet{zemcov} correction.  If we simply averaged the residual signals on the  assumption that the errors for each field were random and uncorrelated, the average would have $\sim4$-$\sigma$ significance.  Only the errors in the total measured sky level for each field is fully random, however.  All components that must be subtracted from the total sky to isolate the dCOB sigma have either fully systematic errors, or errors dominated by systematic uncertainties and only minor random errors.   Estimation of the average residual  COB and dCOB signals and their errors requires care to properly combine such systematic effects with the errors truly random to each field. 

A summary of the origin, amplitude, and type of errors that make up our total error budget is given in Table~\ref{tab:errbudget}. This table presents the error amplitudes as relative errors, $\sigma_x/x$, where $x$ is the intensity, $\lambda I_{\lambda}$, of the indicated sky component. The origins of all of these errors are discussed in detail in previous sections but we present it here to provide a single summary of the error budget as a prelude to our calculation of the final COB and dCOB values. 

\begin{deluxetable*}{llrl}
\tabletypesize{\scriptsize}
\tablecolumns{4}
\tablewidth{\linewidth}
\tablecaption{Relative $\lambda I_{\lambda}$ Error Budget}
\tablehead{
\multicolumn{4}{c}{}\\
\multicolumn{1}{c}{} &
\multicolumn{1}{c}{} &
\multicolumn{1}{c}{Relative Error} &
\multicolumn{1}{c}{}\\
\multicolumn{1}{l}{Flux Component} &
\multicolumn{1}{l}{Origin of Error} &
\multicolumn{1}{c}{($\sigma_{x} / x$)} &
\multicolumn{1}{c}{Type of Error}
}
\startdata
 & & & \\
Total Sky as Observed: 
            & Image to image  variance    & $\sim0.05$ \phantom{gap} & random \\
 & & & \\
\hline
 & & & \\
Faint Stars: 
             & Root(n) in counts at faint end ($V > 22$)     & 0.06 \phantom{gap} & random    \\
             & Same {\tt TRILEGAL} model, different run      & 0.01 \phantom{gap} & random    \\
             & 10\% shift in {\tt TRILEGAL} model parameters & $\sim0.14$ \phantom{gap} & systematic \\
             & Total error:                         & $\sim0.15$ \phantom{gap} & \\
 & & & \\
\hline
 & & & \\
Faint Galaxies: 
       & Root(n) in counts at faint end ($V > 24$) & $<0.01$ \phantom{gap} & random \\
       & Cosmic variance                           & 0.11 \phantom{gap} & random \\
       & Incompleteness corrections \& faint-end slope error & 0.28 \phantom{gap} & systematic \\
       & Total error:                             & 0.30 \phantom{gap} & \\
 & & & \\
\hline
 & & & \\
Scattered Star and Galaxy Light: 
       & Root(n) in cumulative bright star counts         & $<0.01$ \phantom{gap} & random \\
       & Root(n) in cumulative bright galaxy counts       & $<0.01$ \phantom{gap} & random \\
       & PSF uncertainty                       & 0.10 \phantom{gap} & systematic \\
       & Total error                           & 0.10 \phantom{gap} & \\
 & & & \\
\hline
 & & & \\
Diffuse Galactic Light: 
       & Zemcov FIR - Optical DGL slope error & 0.40 \phantom{gap} & systematic \\
       & BD2012 FIR - Optical DGL slope error & 0.23 \phantom{gap} & systematic \\
 & & & \\
\hline
 & & & \\
Scattered Sunlight: 
       & Interplanetary dust                  & 0.00 \phantom{gap} & random \\
       & Scattered light from ACS exhaust$^a$ & 0.00 \phantom{gap} & random \\
       & Sunlight scattered by spacecraft$^b$ & $<0.01$ \phantom{gap} & random \\
 & & & \\
\hline
 & & & \\
Zeropoint Calibration: & Calibration error &   $<$0.01 \phantom{gap}& systematic \\
 & & & \\
\hline
\enddata
\tablecomments{$^a$ Scattered light from the attitude control system exhaust plumes is assumed to be negligible based on gas cooling times and large mean free path lengths. $^b$ Indirect sunlight reflections off of spacecraft into the LORRI aperture are modeled using CAD simulations. The levels of such scattered sunlight entering the LORRI aperture are seen to be negligible in the CAD simulations.}
\label{tab:errbudget}
\end{deluxetable*}

To compute the COB and dCOB values and their uncertainties, we use a Monte Carlo simulation to model all the sources of signal and error. We generate 10,000 realizations of the sky levels in each of our seven fields. For each realization, we generate random normally-distributed values for the sky components. 

The normally-distributed random variables for the six different sky components are generated using the observed mean values and the error values listed in Table~\ref{tab:skyflux}. 
For sky components with statistical (random) uncertainties, an independent random variate is generated for each specific field for each realization. Values for the random error component will thus vary independently for each field and for each realization. 
For sky components with systematic errors, all seven fields are assigned the same Gaussian random variate to generate the given sky intensity component. For example, values for the DGL intensity will vary in the same direction for all seven fields for a given realization. Each realization thus produces a value for the COB and dCOB that accounts for the combined systematic and statistical uncertainties properly.

The final distribution of simulated residual sky levels for all seven fields are combined by weighting the values for each field by the corresponding inverse variance in the residual. We then compute the final mean residual and its 1-$\sigma$ uncertainty from the values in the central 68.3\% probability range derived from the 10,000 weighted residual sky estimates. Using this procedure, our estimates for the final COB residual sky levels (IGL not removed) and for the final dCOB residuals (where the IGL component is subtracted) are given in Table~\ref{tab:cobresults}. The separate estimates of the statistical and systematic errors were made by running a set of simulations with the statistical error components set to zero. We are then able to directly determine what fraction of the total uncertainty is due to the combined systematic errors. For reference, the surface brightness values corresponding to the dCOB intensities given in Table~\ref{tab:cobresults} are 
$27.3^{+0.9}_{-0.5}$ mag~arcsec$^{-2}$ (Zemcov-based DGL) and 
$27.0^{+0.5}_{-0.4}$ mag-arcsec$^{-2}$ (BD2012-based DGL) 
in the AB system computed at the LORRI pivot wavelength of 0.608 ${\rm \mu m}$. 

\begin{deluxetable*}{rcccc|rcccc|l}
\tabletypesize{\footnotesize}
\tablecolumns{11}
\tablewidth{\linewidth}
\tablecaption{Cosmic Optical Background Results}
\tablehead{
\multicolumn{5}{c|}{} & \multicolumn{5}{c|}{} & \multicolumn{1}{c}{}\\
\multicolumn{5}{c|}{COB} & \multicolumn{5}{c|}{Diffuse COB (dCOB)} & \multicolumn{1}{c}{}\\
\multicolumn{1}{c}{} &
\multicolumn{1}{c}{Random} &
\multicolumn{1}{c}{Sys.} &
\multicolumn{1}{c}{Total} &
\multicolumn{1}{c|}{Signif.} &
\multicolumn{1}{c}{} &
\multicolumn{1}{c}{Random} &
\multicolumn{1}{c}{Sys.} &
\multicolumn{1}{c}{Total} &
\multicolumn{1}{c|}{Signif.} &
\multicolumn{1}{c}{Basis of DGL} \\
\multicolumn{1}{c}{Value} &
\multicolumn{1}{c}{Error} &
\multicolumn{1}{c}{Error} &
\multicolumn{1}{c}{Error} &
\multicolumn{1}{c|}{($\sigma$)} &
\multicolumn{1}{c}{Value} &
\multicolumn{1}{c}{Error} &
\multicolumn{1}{c}{Error} &
\multicolumn{1}{c}{Error} &
\multicolumn{1}{c|}{($\sigma$)} &
\multicolumn{1}{c}{Correction}
}
\startdata
 & & & & & & & & & & \\
15.9 & $\pm 1.8$ & $\pm3.7$ & $\pm4.2$ & 3.8 & 
 8.8 & $\pm 1.8$ & $\pm4.5$ & $\pm4.9$ & 1.8 & Zemcov+2017 \\
  & & & & & & & & & & \\
 \hline
  & & & & & & & & & & \\
18.7 & $\pm 1.8$ & $\pm3.3$ & $\pm3.8$ & 4.9 &
11.9 & $\pm 1.8$ & $\pm4.2$ & $\pm4.6$ & 2.6 & BD2012 \\
 & & & & & & & & & & \\
\hline
\enddata
\tablecomments{All COB and dCOB values are in units of nW m$^{-2}$ sr$^{-1}$. Significance is the COB or dCOB value divided by its corresponding total error. }
\label{tab:cobresults}
\end{deluxetable*}
The two estimates of the dCOB signal using the two different DGL calculations are only 1.8-$\sigma$ to 2.6-$\sigma$ significant, but we consider it worthy of further scrutiny. In this section we first review the measurement and decomposition of the sky signal accomplished in the two previous sections, highlighting the major uncertainties with an eye towards where further work might be profitable. We then discuss the extent to which IGL can account for our COB measurement, followed by a review of how it compares to previous attempts to measure the COB.  We finish with a summary of the likelihood that we have recovered evidence for a dCOB in particular.

\subsection{Measuring and Decomposing the Sky}

\subsubsection{Measuring the Total Sky Signal With LORRI}

The first step was to measure the sky signal from LORRI images, which were obtained by a camera optimized for planetary imaging, not low light-level astronomical observations. We examined the basic image reduction procedures and chose to develop an independent reduction pipeline, rather than using the standard New Horizons LORRI pipeline. This included an improved measurement of the image bias level, and an improved understanding of the ``jail-bar'' bias structure, both of which greatly improved the accuracy of the sky determinations. We also discovered a previously unknown background component associated with the activation of the camera prior to an imaging sequence.  Lastly, our ZeroDark65 calibration sequence provided the first direct in flight measurement of the LORRI dark current, showing that it should be  easily characterized as  part of the bias level determination.

In contrast, we did not correct the horizontal streaking in the LORRI background.  While this artifact can readily be removed cosmetically, we did not have a model for how the true zero-level is affected by it in any given image. The streaking pattern does vary from exposure to exposure, however, and we presume that its effect is random and can be reduced by simply averaging over an image data-set. As noted in $\S\ref{sec:bias},$ our tests on in-fight 0s exposures do return a sky level of zero.

The  actual measurement of the sky level in a given LORRI image is done by fitting the peak of the pixel intensity histogram, which provides for robust measurement of the diffuse background.  Care has been taken to account for the highly-structured  histogram associated with low-light levels, but the algorithm used to measure the location of the peak is robust.  The dominant error term is statistical uncertainty in the peak location.  This is reduced by averaging, and modulo any unknown electronic contributions to the background to the images, the final error in the total sky levels should be random.

Lastly, we note that we have benefited from the implementation of long 30s exposures, an operational feature introduced after the Pluto encounter.  Recently, the New Horizons project has enabled the flight system to obtain 65s exposures with LORRI.  While some electronic signals, such as dark current, will scale with exposure time, bias structure should remain constant, and thus be relatively less important in the longer exposures.  No suitable 65s exposures were available for the present work, but exposures of this length should be attractive for future use of LORRI for sky measurements.

\phantom{blank space}

\subsubsection{Starlight}

Starlight contributions come from faint stars within the field and bright stars outside the field.  We benefit from the corpus of modern star catalogs and deep star counts.  At the photometric detection limit of LORRI faint galaxies begin to greatly outnumber faint undetected stars in the field.  Their integral contribution is small and readily provided by TRILEGAL  models.

A more difficult source of starlight to account for is that from bright stars falling outside the LORRI fields.  This term was surprising to us, and was not appreciated until we began exploring the sensitivity of LORRI to scattered light. This term and the DGL term are essentially on parity for the largest contributions to the total observed sky level. As is discussed in \citet{lorri} and \citet{lorri2}, LORRI images obtained within  $90^\circ$ of the Sun always show some level of scattered sunlight.   The amplitude of this scattered light as a function of solar elongation  leads to the large-angle LORRI PSF shown in Figure \ref{fig:LORRIPSF}, which of course must apply to starlight as well.

We ascribe a 10\%\ error to the  effects of the PSF, based on scatter in the amplitude of sunlight with azimuth at any radial elongation. This error is effectively systematic, as it affects all fields similarly.  In contrast, knowledge of the stars that contribute to any given field, as provided by the  Tycho and {\it Gaia} catalogs, is highly accurate. As with the treatment of galaxies, we have  used color measurements to provide for accurate application of the LORRI response to any  given star.

We consider it unlikely that the contribution of bright stars to the total sky level has been underestimated. It would take nearly a 100\%\ error in the amplitude of the PSF to explain the  residual COB signal. A LORRI calibration program that attempts to image the scattering wings of a zero-magnitude star over a range of small elongations might be useful to tie down uncertainties in the PSF.

In passing, we note that the scattered light component from bright {\it galaxies} lying outside the fields is negligible (see Table~\ref{tab:skyflux}). The number counts of galaxies fall off rapidly with increasing apparent brightness, dropping by a factor of 4 for every 1 magnitude decline when $V \leq 19$.

\subsubsection{Diffuse Galactic Light from the Milky Way}

The DGL component contributes significantly to the total sky level, and the error in the slope of the  conversion between $100~{\rm\mu m}$ flux from the infrared cirrus and the optical flux of scattered Milky Way light is the largest contribution to the total error in the COB signal. Our knowledge of the DGL is not fully satisfactory.  The \citet{zemcov} estimator has large errors, which are based on theoretical uncertainties in its derivation.  The  DGL estimator based on the \citet{BD2012} observations of DGL spectra is of markedly smaller amplitude and has smaller formal errors, but depends on a poorly determined correction for a bias in the analysis methodology. We have carried results from both DGL estimators in parallel to gauge the overall likely spread in the COB and dCOB levels due to uncertainty in the DGL correction.

\subsection{The Integrated Light from Galaxies}\label{sec:igldiscuss}

 The IGL -- integrated light from galaxies falling within the LORRI fields but that are too faint to be detected individually -- nominally accounts for $\sim50\%$ of our total COB signal. The difference between our measurements of the COB and the IGL is significant at the $\sim2\sigma$ level. At this level of significance we indeed cannot falsify the hypothesis that our COB measure is solely explained by the IGL without recourse to an additional dCOB component. Conversely, it may be possible that our COB measure admits evidence for a visual-light source not explained by the present inventory of galaxies.  We explore these competing hypotheses in this and the next subsection.
 
\begin{figure*}[htbp]
\centering
\includegraphics[keepaspectratio,width=6 in]{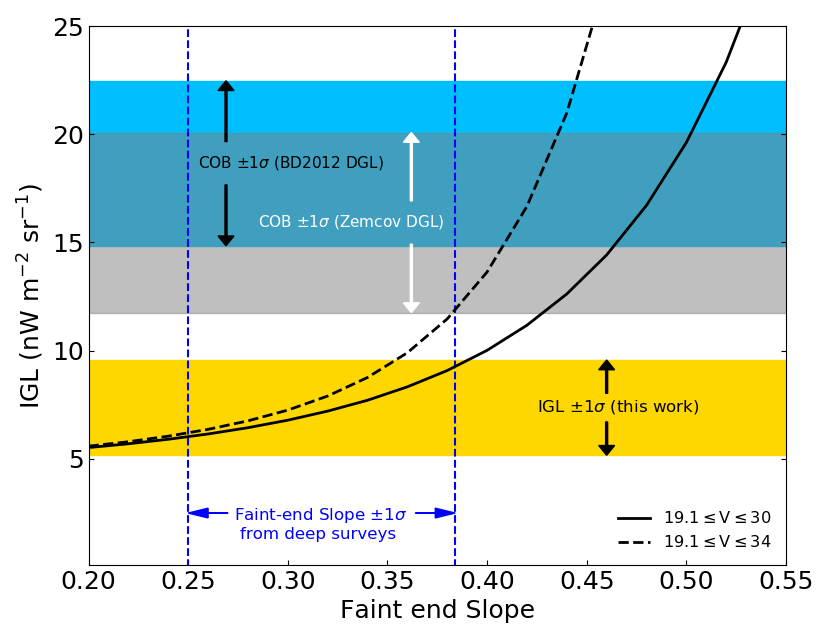}
\caption{The dependence of the derived IGL on the faint end slope of the relationship between the differential number counts of galaxies and their apparent V magnitude. The solid and dasked black curves show the trends for integration of the galaxy number counts performed over the range $19.1 \leq V \leq 30$ and $19.1 \leq V \leq 34$, respectively. The vertical blue lines show the $\pm 1\sigma$ observational constraints on the faint end slope from various deep galaxy surveys. The blue and grey filled regions show our $\pm 1\sigma$ limits on the COB using DGL estimates based on \cite{BD2012} and \cite{zemcov}, respectively. The gold filled region shows our $\pm 1\sigma$ constraints on the IGL from undetected faint galaxies as described in $\S\ref{ssec:fgal}$.}
\label{fig:faintendslope}
\end{figure*}

 In attempting to understand the IGL component we again benefit from an immense corpus of work done on galaxy counts in multiple optical bandpasses that allows us to compute constraints, including extrapolation to fainter fluxes ($\ge 28$ mag), where direct observational constraints are only now just becoming available. The two key factors in setting the value of the derived IGL are the normalization and the faint end slope of the relationship between the logarithm of the differential galaxy counts as a function of apparent magnitude. Figure~\ref{fig:faintendslope} shows how our derived IGL would change as a function of the faint end slope assumed for the V-band galaxy count - magnitude relation. The trends in the figure were derived by setting the normalization of the galaxy number counts to match that observed at V = 23 (where the incompleteness corrections are negligible) but then allow the slope to vary from $0.2 - 0.6$ at fainter magnitudes. The IGL, integrated down to V=30 mag, would reach the observed COB level if the faint end slope were $\sim 0.48$. If the IGL is derived from an integration down to V=34 mag, the faint end slope would need to be at least $0.42$ to match the observed COB level. Observational constraints on the faint end slope are denoted by the vertical blue lines in Figure~\ref{fig:faintendslope}. While such steep faint end slopes are not conclusively ruled out, the observations and theoretical predictions strongly favor shallower slopes, with values $< 0.38$. At $V > 27$ in the deepest existing galaxy surveys, the number count - magnitude relations typically have slopes $< 0.25$. 

Alternatively, one can ask what normalization shift in the number counts would be required to have the IGL explain the full COB signal. We estimate that, for the observed faint end slope of $\sim 0.32$, the differential galaxy counts would have to be underestimated by a factor of $2.0 \pm 0.5$. Since the differential galaxy count data we use have already been corrected for incompleteness, such a normalization shift would imply that existing deep galaxy surveys are systematically missing about half of the actual galaxies with $V < 30$. 

\citet{conselice} performed comprehensive study of how many galaxies could occupy the visible universe. They estimate that there may be up to 10 times the number of galaxies than currently counted in existing deep surveys. This difference is driven by their extrapolation of the stellar mass function down to $10^6$ M$_{\odot}$ and thus most of those galaxies would have apparent magnitudes fainter than 30 mag. They would nonetheless contribute to the total COB. Indeed,  \citet{conselice} provide an accompanying estimate of the IGL that is $5.8 \times 10^{-9}$ erg s$^{-1}$ cm$^{-2}$ \AA$^{-1}$ sr$^{-1}$ for an integration down to 30 mag, which, at a wavelength of 0.608$~\mu$m, corresponds to $35 {\rm ~nW~m^{-2}~sr^{-1}}$. Their estimate of the IGL is comparable to our total sky level even before correction for the known foregrounds. Our observational limits on the COB would rule out that specific value for the IGL at very high significance but as their integrated number counts appear to have about a 35\% uncertainty, their IGL estimate may yet be compatible with our observational constraint. The difference between the \citet{conselice} estimate of the IGL and our observations might, however, suggest that extending the Schechter form for the galaxy mass function down to $10^6~{\rm M}_{\odot}$ is not fully warranted.

An independent assessment of the IGL is provided by analysis of data from the High Energy Stereoscopic System (\citet{hess}). H.E.S.S. detected very high energy ($>100$GeV) $\gamma$-rays emitted by a sample of blazars over a $0.03<z<0.19$ redshift range. At these energies, the space density of optical photons can provide a significant cross section for the production of electron-positron pairs, thus attenuating the $\gamma$-ray flux. While this is an indirect constraint on the COB flux density, it truly probes the extragalactic optical photon density, and is not vulnerable to solar system zodiacal light or even Milky Way optical foreground emission.  Furthermore, observing background sources at a range of redshifts verifies that the  strength of the attenuation increases with path-length. The integration of our galaxy counts over the magnitude range [12, 30] agrees quite well with the H.E.S.S. constraints, as shown in Figure~\ref{fig:cob}. If the COB is due solely to the collective light from faint galaxies, then the $2\sigma$ difference between our derived IGL and our estimate of the COB would imply a factor of $\sim2$ under-count of galaxies even at apparent magnitudes well within the grasp of current telescopes.

Finally, we note that although we needed to estimate how the SEDs of galaxies vary with galaxy magnitude to account for the wide LORRI bandpass, the available observations provide reasonably good constraints on that variation and, in the end, any uncertainty in this step of the analysis is not as significant as uncertainties in the galaxy number counts discussed above. 

\subsection{Comparison to Previous COB Measurements}\label{sec:comparison}

\begin{figure*}[htbp]
\centering
\includegraphics[keepaspectratio,width=6.0 in]{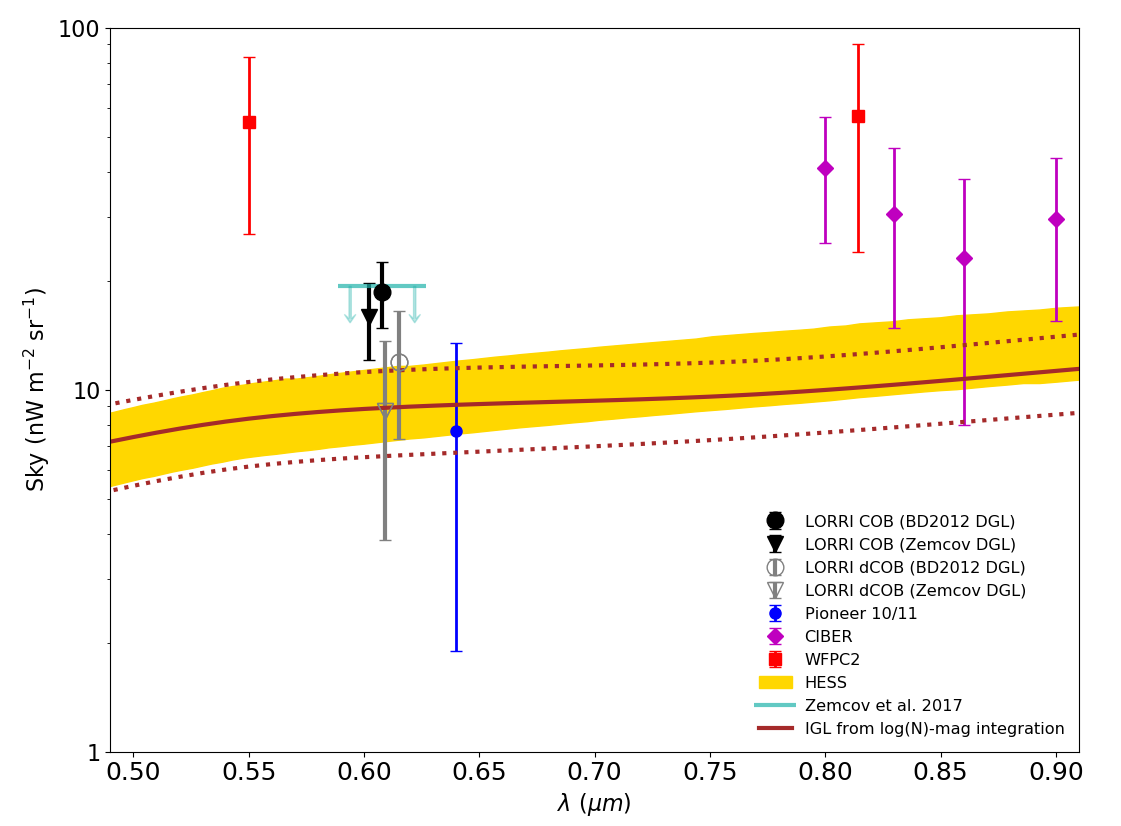}
\caption{The present LORRI COB and dCOB signals are compared to previous COB measures. The solid black circle and triangle symbols are the LORRI COB signal using the BD2012 and Zemcov DGL estimates, respectively. The open grey symbol are the corresponding COB results corrected for the IGL component, producing estimates of a diffuse COB (dCOB) component. (All 4 of these points are slightly horizontally shifted with respect to one another for clarity). Note that while the LORRI points are plotted at their appropriate pivot wavelength, the LORRI bandwidth is larger than the range of the figure (see Figure \ref{fig:lorribandpass}). The green horizontal line with the down arrows is the \citet{zemcov}  2-$\sigma$ COB upper limit. Other COB results are Pioneer 10/11 \citep{pioneer}, CIBER \citep{ciber}, WFPC2 \citep{wfpc2}, and H.E.S.S. \citep{hess}. IGL is the range of flux contributed by integrating galaxy number counts over the mag range [12,30] for the UBVRIz passbands (note that, in this instance, the integration over LORRI passband is not required).}
\label{fig:cob}
\end{figure*}

The present COB measurement appears to be in good agreement with previous measurements, but probes an interesting part of parameter space that should allow for new constraints on a true diffuse background of cosmological origin.   Figure \ref{fig:cob} shows our measurements in the context of previous COB measurements relevant to the LORRI band-pass.   In reviewing the literature, we note that our emphasis on detecting a dCOB signature differs  from most studies, which prefer to measure the total COB and compare it to the IGL inferred from deep galaxy  counts. 

For reference, Figure \ref{fig:cob} also includes our estimate of IGL as a function of wavelength based on the galaxy counts presented  in $\S \ref{ssec:fgal}.$  The IGL shown here is independent of the LORRI band-pass.  This an important caveat when evaluating Figure \ref{fig:cob}.  The actual IGL value we subtracted from our measures is an integral over the LORRI band-pass, and thus is slightly different than the IGL presented in the figure, which was done here for comparison to all previous COB observations.  The IGL trace shown is derived by integrating the multi-power law fits to the galaxy number counts (see Figure~\ref{fig:galNmag}) in the UBVRIz bands over the magnitude interval [12,30] using equation~\ref{eq:sbint} and converting the resulting surface brightness to intensity ($\lambda I_{\lambda}$) at the effective wavelengths of each pass-band. The red curve in Figure \ref{fig:cob} is a spline fit to these intensities. The position of the COB measurements relative to the IGL contribution indicates the contribution of a possible diffuse component.

Our present observations are obviously most closely related to the earlier work done using LORRI by \citet{zemcov}.  Our respective results are compatible, even though there are substantial differences in in the image set, data processing, and analysis methodologies.  We based  our study on 195 30s exposures spread over seven fields, while Zemcov et al.\ were only able to identify four single 10s archival images of four different fields that were suitable for their use; the subsequent improvement in material suitable for deep sky measurements is due to the DKBO and zodiacal light programs initiated after their analysis was published. Our program also benefited from improved understanding of the low light-level performance of LORRI, and knowledge of its sensitivity to scattered light.  This said, Zemcov et al.\ represented their measures conservatively as a 2-$\sigma$ upper limit on the EBL of  $19.3{\rm ~nW~m^{-2}~sr^{-1}},$  which readily accommodates our COB and dCOB measurements.   Zemcov et al.\ also produced an EBL estimate of $4.7\pm10.3{\rm ~nW~m^{-2}~sr^{-1}},$ which is not significantly different  from our result (not shown in the figure).

\citet{wfpc2} used {\it HST}/WFPC2 images  to measure a COB level of $55\pm 28{\rm ~nW~m^{-2}~sr^{-1}}$ at $0.55~\mu{\rm m}$ and $57\pm 33{\rm ~nW~m^{-2}~sr^{-1}}$ at $0.81~\mu{\rm m}.$ While these flux levels are markedly higher that our COB result, with their large  error bars this difference is not highly significant.  We note that our total sky level, $33.2\pm0.5{\rm~nW~m^{-2}~sr^{-1}},$  {\it prior to any corrections} is itself only $\sim60\%$ of the \citet{wfpc2} COB signal.

In contrast, the COB measurement of $7.5\pm 5.8{\rm ~nW~m^{-2}~sr^{-1}}$ at $0.64~\mu{\rm m}$ provided by Pioneers 10 and 11 photometry  \citep{pioneer},  is $\sim50\% $ of our value,  but again has low significance.  This work is of particular interest, however, as the photometry was obtained when the spacecraft were 3.2 to 5.2 AU from the Sun, allowing the authors to claim that the effects of zodiacal light were negligible.  More recently, however, \citet{matsumoto} re-reduced the Pioneer 10 data and claimed that instrumental artifacts prevent accurate measurement of the absolute sky level. 

\citet{ciber} obtained low-resolution $0.8~\mu{\rm m}\leq\lambda\leq1.8~\mu{\rm m}$ NIR spectra of the sky using the Cosmic Infrared Background Experiment (CIBER) rocket instrument. They argued that their COB measurements were significantly above the expected IGL contribution, and had a spectral energy distribution that could not be matched by improperly subtracted ZL.  While the center  of our  band-pass is well to the blue of their spectral range, it does overlap with the CIBER observations, and a dCOB signal at $<0.9~\mu{\rm m}$ would contribute to our sky.  The dCOB level that we see is compatible with their  measurements.

The \citet{hess} COB constraints are represented as the gold-colored band in Figure \ref{fig:cob}.  The H.E.S.S. collaboration emphasized the close concordance of their results with the known IGL component, which is evident in our figure. At the same time, it is also clear that the H.E.S.S. EBL measurements would allow a dCOB component, consistent with our result, particularly in the NIR.

The previous COB measurements discussed so far are estimates of an overall constant background level.  A different technique is to look for angular fluctuations in the background to assess the light associated with sources fainter than the photometric detection limit.  An advantage to this approach is that it rejects ZL, which is assumed to be smooth on fine scales. \citet{zem14} analyzed the NIR CIBER images and argued for a dCOB component (in our parlance) of $7.0^{+4.0}_{-3.5}{\rm ~nW~m^{-2}~sr^{-1}}$ at $1.1~\mu{\rm m},$ which is consistent with the present results.  They argued that this might be evidence for intra-halo light (IHL), which would be a relatively local component of stars tidally ripped out of merging galaxies \citep{cooray12}. \citet{mat19} analyzing optical-band fluctuations in the Hubble Extremely Deep Survey \citep{xdf} also argue for a significant dCOB, which would be generated by faint compact objects at $z\lesssim0.1.$ A key argument for sources at low redshift is accommodating the very high energy $\gamma$-ray opacity limits on COB components at $z>>0.1.$

We finish this discussion on the possibility of a dCOB by returning to our discussion of \citet{conselice} in the previous subsection.  The central thesis of that work is that existing galaxy counts fall an order of magnitude short of accounting for the total visible light output of stars integrated over the history of the Universe. While \citet{conselice} in fact predict a COB substantially in {\it excess} of what our observations and the H.E.S.S. results can accommodate, their conclusion that the present view afforded by galaxy counts may not be all there is to the story remains as an observational challenge.


\acknowledgments
We thank NASA for their funding and continued support of the New Horizons mission.  The data presented here was obtained during the Kuiper Extended Mission of New Horizons. This work has made use of data from the European Space Agency (ESA) mission {\it Gaia} (\url{https://www.cosmos.esa.int/gaia}), processed by the {\it Gaia} Data Processing and Analysis Consortium (DPAC, \url{https://www.cosmos.esa.int/web/gaia/dpac/consortium}). Funding for the DPAC has been provided by national institutions, in particular the institutions participating in the {\it Gaia} Multilateral Agreement. We sincerely thank Dr. Caroline Bot for providing over 5 million IRIS $100\ \mu$m flux measurements for the high-galactic latitude regions of the sky and for helpful discussions. We thank Drs. Bruce Draine and Tim Brandt for several very useful conversations about the physics of scattering of optical starlight by dust in the Milky Way. We thank Dr. Michele Trenti for providing us with their cosmic variance calculation software. We thank the referee for a thorough reading of the manuscript and a number of useful suggestions.

\software{astropy \citep{2013A&A...558A..33A, 2018AJ....156..123A}, matplotlib \citep{matplotlib}, vista \citep{vista}}

{}

\end{document}